\documentclass[10pt,twocolumn, cleanhead, cleanfoot]{asme2e}

\usepackage{empheq, amssymb}

\usepackage{lmodern}
\usepackage{amsmath}
\usepackage{amsfonts}
\usepackage{bm} 
\usepackage{txfonts}
\usepackage{tabularx}
\usepackage[T1]{fontenc}
\usepackage[utf8]{inputenc} 
\usepackage[dvipsnames]{xcolor} 
\usepackage[american]{babel} 
\usepackage{graphicx}
\usepackage{epstopdf}
\usepackage{booktabs}
\usepackage[noadjust]{cite} 
\usepackage{ltxcmds}
\usepackage{mathtools}
\usepackage{subcaption}
\usepackage{multirow} 
\usepackage{blindtext}
\PassOptionsToPackage{hyphens}{url}
\usepackage{hyperref}
\usepackage{enumitem}
\usepackage{ifthen}
\usepackage{etoolbox}
\usepackage{cancel}
\usepackage{empheq}
\usepackage{fontawesome}
\usepackage{soul}
\usepackage{flushend}
\usepackage{textcomp}
\usepackage{lipsum}

\definecolor{light-gray}{gray}{0.4}
\definecolor{box-gray}{gray}{1}

\hypersetup{
    unicode=false,          
    pdftoolbar=true,        
    pdfmenubar=true,        
    pdffitwindow=false,     
    pdfstartview={FitV},    
    pdftitle={A General framework for supporting economic feasibility of generator and storage energy systems through capacity and dispatch optimization},    
    pdfauthor={Saeed Azad, Ziraddin Gulumjani, Daniel R. Herber},     
    pdfsubject={},   
    pdfkeywords = {}, 
    pdfnewwindow=true,      
    colorlinks=true,
    allcolors=blue
}

\usepackage{nomencl}

\renewcommand\nomgroup[1]{%
  \item[\bfseries
  \ifstrequal{#1}{V}{ Select Variables}{%
  \ifstrequal{#1}{B}{ Subscripts}{%
  \ifstrequal{#1}{P}{ Notation}{%
  \ifstrequal{#1}{A}{ Acronyms}{}}}}]
}

\makenomenclature



\usepackage{dblfloatfix}
\usepackage{float}

\definecolor{block-gray}{gray}{0.95}

\usepackage[framemethod=TikZ]{mdframed}

\usepackage{xpatch}

\makeatletter
\xpatchcmd{\endmdframed}
  {\aftergroup\endmdf@trivlist\color@endgroup}
  {\endmdf@trivlist\color@endgroup\@doendpe}
  {}{}
\makeatother

\usepackage{fontawesome}

\usepackage{titlesec}

\usepackage{hyperref}

\newcommand{\doi}[1]{{doi:~\href{http://doi.org/#1}{#1}}\rmFullStop}

\newcommand*{\rmFullStop}{\rmifnextchar{.}{}{}}

\makeatletter
\newcommand{\rmifnextchar}[3]{%
  \begingroup
  \ltx@LocToksA{\endgroup#2}%
  \ltx@LocToksB{\endgroup#3}%
  \ltx@ifnextchar{#1}{%
    \def\next{\the\ltx@LocToksA}%
    \afterassignment\next
    \let\scratch= %
  }{%
    \the\ltx@LocToksB
  }%
}
\makeatother

\usepackage{colortbl}

\definecolor{light-gray}{gray}{0.6}

\newcommand{\xsection}[1]{\section[#1]{\MakeUppercase{#1}}}

\usepackage{algorithm2e}
\usepackage{xcolor}

\SetAlFnt{\footnotesize}

\SetAlCapSty{xAlCapSty}

\definecolor{commentcolor}{HTML}{1E4D2B}

\SetCommentSty{xCommentSty}

\SetNlSty{mynlfont}{}{} 

\LinesNumbered

\SetSideCommentRight

\DontPrintSemicolon

\RestyleAlgo{algoruled}

\usepackage{algorithm2e}
\usepackage{etoolbox}
\usepackage{xstring}
\usepackage{calc}
 
\newlength{\xalgowidth}
\newlength{\xalgoremainder}
\newlength{\xindentwidth}

\newenvironment{vAlgorithm*}[3][]{
  \setlength{\xalgowidth}{#2} 
  \setlength{\xindentwidth}{#3} 
  \setlength{\xalgoremainder}{\textwidth-\xalgowidth} 
  \SetCustomAlgoRuledWidth{\xalgowidth} 
  \IncMargin{\xindentwidth}
  \begin{algorithm*}[#1]
}
{
  \end{algorithm*} 
  \DecMargin{\xindentwidth}
}

{
  \end{algorithm} 
  \DecMargin{\xindentwidth}
}

\makeatletter
\patchcmd{\@algocf@start}{%
\begin{lrbox}{\algocf@algobox}%
}{%
\rule{0.5\xalgoremainder}{\z@}
\begin{lrbox}{\algocf@algobox}%
\begin{minipage}{\xalgowidth}%
}{}{}
\patchcmd{\@algocf@finish}{%
\end{lrbox}%
}{%
\end{minipage}%
\end{lrbox}%
}{}{}
\makeatother

\usepackage{accents}

\usepackage[most]{tcolorbox} 
\definecolor{block-gray}{gray}{0.95}

\AtBeginDocument{%
 \abovedisplayskip=4pt plus 1pt minus 1pt
 \abovedisplayshortskip=0pt plus 3pt
 \belowdisplayskip=4pt plus 1pt minus 1pt
 \belowdisplayshortskip=0pt plus 3pt
}

\newtcolorbox{xreviewer}{%
	empty,
    borderline west = {4pt}{0pt}{gray},
    boxrule = 0pt,
    boxsep = 0pt,
    breakable,
    colback = block-gray,
    enhanced,
    frame hidden,
    left skip = 0pt,
    notitle,
    parbox = false,
    sharp corners,
}
\newtcolorbox{xresponse}{%
	empty,
    boxsep = 0pt,
    breakable,
    frame hidden,
    notitle,
    parbox = false,
}

\newtcolorbox{xchange}{%
    enhanced jigsaw,
    breakable,
    boxrule = 1pt,
    boxsep = 0pt,
    colframe = black,
    colback = white,
    notitle,
    parbox = false,
    arc = 0pt,
    outer arc = 0pt,
    after skip = 10pt plus 2pt,
    before upper={\color{violet}},
    before upper=\color{violet},  
    before upper*=\color{violet}, 
    every upper part/.style={before=\color{violet}}, 
    use color stack,
}

\usepackage{catchfilebetweentags}
\makeatletter
\def\CatchFBT@Fin@l#1[#2]{%
   \begingroup
      \makeatletter #2%
      \scantokens\expandafter{%
         \expandafter\CatchFBT@tok\expandafter{\the\CatchFBT@tok}}%
      \CatchFBT@IsAToken{#1}
         {\global#1\expandafter{\the\CatchFBT@tok}}
         {\xdef#1{\the\CatchFBT@tok}}%
      \ifx\CatchFBT@tok#1\else\global\CatchFBT@tok{}\fi
   \endgroup
}
\makeatother

\newcommand{\parm}{\mathord{\color{black!33}\bullet}}%
\usepackage{cancel}

\definecolor{Pcolor}{HTML}{800000}
\definecolor{Ecolor}{HTML}{084f09}
\definecolor{Tcolor}{HTML}{4e79a6}
\definecolor{ZTcolor}{HTML}{8a3c9e}
\definecolor{Mycolor}{HTML}{ee5622}
\definecolor{PRcolor}{HTML}{E3B505} 

\definecolor{CRcolor}{HTML}{006955}
\definecolor{GScolor}{HTML}{21447a}
\definecolor{Neutralcolor}{HTML}{000000}
\definecolor{ORcolor}{HTML}{A90432}

\newcommand{\PC}[1]{\textcolor{Pcolor}{#1}}
\newcommand{\EC}[1]{\textcolor{Ecolor}{#1}}
\newcommand{\TC}[1]{\textcolor{Tcolor}{#1}}

\newcommand{\CR}[1]{\textcolor{CRcolor}{#1}}
\newcommand{\GS}[1]{\textcolor{GScolor}{#1}}
\newcommand{\OR}[1]{\textcolor{ORcolor}{#1}}

\usepackage{units}
\usepackage{nicematrix}

\newcommand{\ToolName}{ECOGEN-CCD}

\usepackage{accents}
\usepackage{units}
\newcolumntype{s}{>{\columncolor[HTML]{F5F5F5}} c}
\makeatletter
\newcommand*{\rom}[1]{\expandafter\@slowromancap\romannumeral #1@}
\makeatother

\usepackage{tikz}
\newcommand*\circled[1]{\tikz[baseline=(char.base)]{
            \node[shape=circle,draw,inner sep=0.6pt] (char) {#1};}}

\usepackage{makecell}

\usepackage{tikz}
\usetikzlibrary{matrix,fit} 
\usetikzlibrary{matrix}
\usetikzlibrary{graphs}
\usepackage{pgfkeys} 
\usetikzlibrary{positioning}
\usepackage{lipsum,adjustbox}
\usetikzlibrary{arrows.meta}

\usepackage{overarrows}

\NewOverArrowCommand{overrightharpoon}{end=\rightharpoonup}
\NewOverArrowCommand{overleftharpoon}{end=\leftharpoonup}

\newcommand{\overleftharpoonsub}[2]{\overleftharpoon{#1}_{\hspace{-0.22em}#2}}
\newcommand{\overrightharpoonsub}[2]{\overrightharpoon{#1}_{\hspace{-0.22em}#2}}

\usepackage[displaymath,switch]{lineno}

\newcommand*\patchAmsMathEnvironmentForLineno[1]{%
  \expandafter\let\csname old#1\expandafter\endcsname\csname #1\endcsname
  \expandafter\let\csname oldend#1\expandafter\endcsname\csname end#1\endcsname
  \renewenvironment{#1}%
     {\linenomath\csname old#1\endcsname}%
     {\csname oldend#1\endcsname\endlinenomath}}%
\newcommand*\patchBothAmsMathEnvironmentsForLineno[1]{%
  \patchAmsMathEnvironmentForLineno{#1}%
  \patchAmsMathEnvironmentForLineno{#1*}}%
\AtBeginDocument{%
\patchBothAmsMathEnvironmentsForLineno{equation}%
\patchBothAmsMathEnvironmentsForLineno{align}%
\patchBothAmsMathEnvironmentsForLineno{flalign}%
\patchBothAmsMathEnvironmentsForLineno{alignat}%
\patchBothAmsMathEnvironmentsForLineno{gather}%
\patchBothAmsMathEnvironmentsForLineno{multline}%
}

\title{A General framework for supporting economic feasibility of generator and storage energy systems through control co-design optimization}

\author{Saeed~Azad\thanks{Corresponding author, \texttt{\href{mailto:s-azad@onu.edu}{s-azad@onu.edu}}} 
\affiliation{
Visiting Assistant Professor\\
Mechanical Engineering Department \\
Ohio Northern University\\
Ada, OH 45810 \\
~\texttt{\href{mailto:s-azad@onu.edu}{s-azad@onu.edu}}\\ 
}
}

\author{Ziraddin~Gulumjanli
\affiliation{
Graduate Student\\
Department of Systems Engineering \\
Colorado State University\\
Fort Collins, CO 80523 \\
~\texttt{\href{mailto:ziraddin.gulumjanli@colostate.edu}{ziraddin.gulumjanli@colostate.edu}}\\ 
}
}

\author{Daniel~R.~Herber
\affiliation{
Associate Professor\\
Department of Systems Engineering \\
Colorado State University \\
Fort Collins, CO 80523 \\
~\texttt{\href{mailto:daniel.herber@colostate.edu}{daniel.herber@colostate.edu}}
}
}

\usepackage{amsmath}

\begin{document}
 \setlength{\parskip}{0pt}
 \setlength{\parsep}{0pt}
 \setlength{\headsep}{0pt}

\setlength{\topsep}{0pt}

\abovedisplayshortskip=3pt
\belowdisplayshortskip=3pt
\abovedisplayskip=3pt
\belowdisplayskip=3pt

\titlespacing*{\section}{0pt}{18pt plus 1pt minus 1pt}{3pt plus 0.5pt minus 0.5pt}

\titlespacing*{\subsection}{0pt}{9pt plus 1pt minus 0.5pt}{1pt plus 0.5pt minus 0.5pt}

\titlespacing*{\subsubsection}{0pt}{9pt plus 1pt minus 0.5pt}{1pt plus 0.5pt minus 0.5pt}

\maketitle

\begin{abstract}\noindent
\textit{
Hybrid energy systems (HES) entail the integration of various electricity-generating technologies (such as natural gas, wind, nuclear, etc.) and storage systems (such as thermal, battery electric, hydrogen, etc.).
Such integration efforts, particularly when focused on identifying superior configurations, face challenges in determining the optimal storage sizing and dynamic behavior and necessitates an adaptable and efficient evaluation framework.
Driven by such needs in both early system concept development and retrofit efforts, this work outlines a versatile computational framework for efficiently assessing the net present value of various integrated generator/storage technologies with a general optimization model.
The subsystems' fundamental dynamics are defined, with a particular emphasis on balancing critical physical and economic domains to enable optimal decision-making in the context of control co-design (CCD).
In its presented form, the framework formulates a linear dynamic optimization problem that can be efficiently solved through a direct transcription approach.
The CCD optimization problem of an HES for $30$ years with an hourly mesh can be solved in less than $1800~[\unit{s}]$, depending on the case study.
Three case studies focusing on natural gas with thermal storage and carbon capture, wind energy with battery storage, and nuclear with hydrogen are selected to demonstrate the framework's capabilities in formulating a wide range of HES problems in the context of CCD and dynamic optimization, highlighting its value in facilitating the techno-economic assessment of various HES configurations. 
}
\end{abstract}

\vspace{1ex}

\xsection{Introduction}\label{sec:introduction}

The increasing energy demand resulting from various technological advancements (such as artificial intelligence \cite{AI} and vehicle electrification), population growth, and changing energy consumption patterns has significantly affected various aspects of the energy market.
This increase coincides with environmental policies that dictate more stringent requirements on carbon emissions \cite{US_2021}, driving an increase in renewable energy generation.

Due to their intermittent, unpredictable, and variable nature, renewable energy generation faces challenges in grid stability and controllability \cite{Mario_2024, NREL_2024}. 
Moreover, from the perspective of plant operators, it is increasingly challenging for new single-technology projects (renewable and/or non-renewable) to remain economically profitable \cite{NREL_2024}.
Hybrid energy systems (HES) are a promising concept that integrates multiple generation, storage, and conversion technologies to provide energy or commodity products (such as hydrogen) \cite{Wu2016, Mario_2024, murphy_taxonomy_2021}.
If properly designed and planned \cite{sinha2014review}, hybridization enables an increase in grid stability, control authority, flexibility, and robustness \cite{Arent2021, Zhang2022}, and even profitability of utility companies \cite{Hancock2022, Zhang2022}.

Informed by market restrictions, policy requirements, and energy economics, HES can be configured in a variety of ways, each with different generation and/or storage technologies.
Some examples include photovoltaics (PV) + wind \cite{khare2016solar}, solar + storage \cite{osti_2447818}, PV + wind + hydropower + pumped-storage \cite{xu2020optimized}, and nuclear + wind + hydrogen storage \cite{Zhang2022}.
Far from simply adding an independent subsystem, effective HES designs must ensure that interactions among energy technologies are synergistic, thereby economically benefiting the entire system \cite{Mario_2024}.

To be economically beneficial, the flexibility added through hybridization must be sufficient to overcome the relevant costs, including capital and operational costs of the technology over its lifetime.
A techno-economic assessment, presented here as a net present value (NPV) objective, is often at the core of analyzing how the integration of new units (such as generator and storage) and functions (such as carbon capture and storage (CCS), district heating/cooling, etc.) economically affects the entire system.
It is also central to any retrofit design efforts within a structured energy system.
Such economic objectives are particularly well-suited for use with novel design frameworks, such as control co-design (CCD).

CCD is a system-level optimization strategy in which both the physical attributes (such as capacity) and control (such as dispatch, charge, and discharge) are concurrently optimized to enhance system performance \cite{GarciaSanz2019, Azad2023, Herber2017e}.
By assessing the synergy or dysfunction among various energy system technologies, CCD helps guide the early-stage decision-making in the design of HESs \cite{Mario_2024}.
CCD approaches been explored for a natural gas combined cycle (NGCC) power plant with thermal energy storage (TES) and CCS \cite{Vercellino2022}, light water reactor (LWR) with various TES technologies \cite{McDowell2021}, and a nuclear power plant (NPP) with hydrogen production \cite{SotoGonzalez2022}.

\
While co-designing HES offers numerous advantages, the computational tools currently available often require extensive domain knowledge of specific energy technologies to develop high-fidelity, nonlinear models and frequently custom implementations. 
The resulting dynamic optimization problem, therefore, is often computationally expensive, making early-stage conceptual analysis of various HES configurations challenging.
One example is the Holistic Energy Resource Optimization Network (HERON) toolbox \cite{doecode_45038}.
In a case study in Ref.~\cite{Li2021}, instead of solving the problem for an entire year, the authors considered a $24$-hour horizon and solved the problem $365$ times to reduce complexity and computational cost.
Given that effective early-stage decision-making, ideally requires only a few assumptions and adjustments beyond the core techno-economic ones (considering that specific system components, characteristics, and performance criteria are often unknown and will be determined in later steps), an opportunity exists to leverage subsystems' basic dynamics to develop an efficient and versatile framework to assess various system configurations and scenarios for hybrid generator and storage energy systems.

Therefore, this article presents the development and demonstration of an open-source computational framework in \texttt{MATLAB} for assessing the 
\textit{economic feasibility of generator and storage systems through CCD}, which we refer to as \ToolName.
Empowered by subsystems' fundamental dynamics and an NPV objective function, \ToolName{} enables the inclusion of various generation technologies, functions (such as CCS), and storage types to help support optimal decision-making in exploring a wide range of HES configurations. 
While easily configurable for assessing different HES concepts, \ToolName~does not require extensive knowledge of specific energy technologies.
Instead, the user is only required to provide basic technical and economic parameters to define the problem (see Fig.~\ref{fig:OptModel} for an overview of the needed parameters).%

Once formulated, the HES CCD optimization problem can be implemented and solved using the open-source \texttt{MATLAB} software DTQP \cite{Herber2017e, DTQP}, which uses direct transcription with an automated optimization problem generation procedure.
In its presented form, the framework formulates a linear optimization problem that can be efficiently solved with appropriate solvers.  
We note that compared with the tools utilized in Refs.~\cite{GenX} and \cite{doecode_45038}, \ToolName{} emphasizes the design and operation of an individual plant, consisting of a collection of generators, storage units, and functions. 
\ToolName{} is an open-source tool that is made publicly available in Ref.~\cite{code}.

\begin{figure*}[t]
    \centering
    \includegraphics[scale=1]{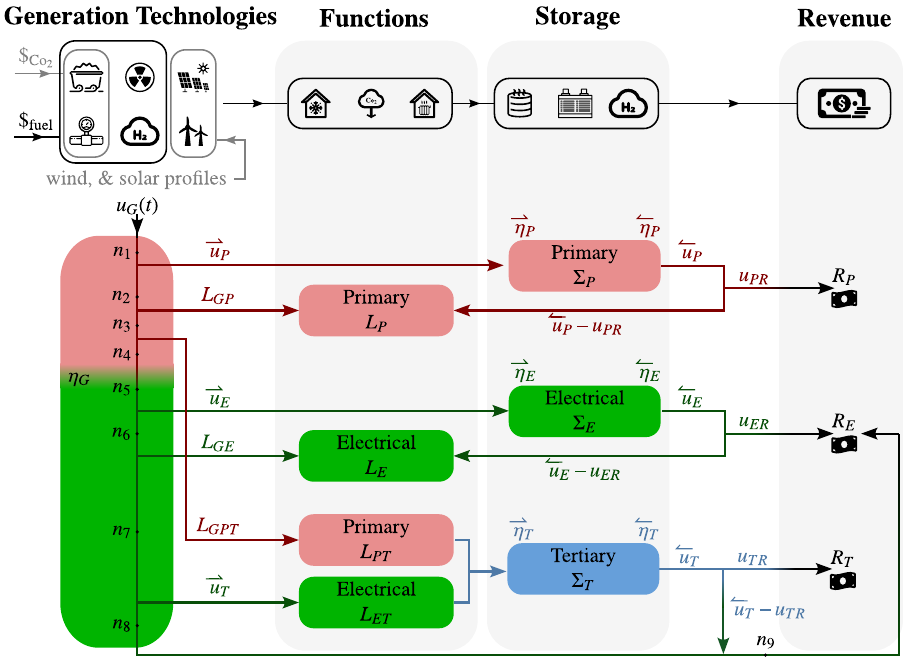}
    \caption{Illustration of an HES architecture with a collection of homogeneous electric power generators and three types of storage systems. Primary (such as thermal), electrical (such as battery), and tertiary (such as hydrogen) storage is shown in \PC{red}, \EC{green}, and \TC{blue}, respectively. The charge and discharge signals, along with their associated efficiencies, are described by $\protect\overrightharpoon{\parm}$ and $\protect\overleftharpoon{\parm}$, respectively.} 
    \label{fig:IES}
\end{figure*}

The remainder of the article is organized in the following manner:
Sec.~\ref{sec:section2} starts by describing some motivations for the proposed framework and then discusses framework architecture, problem elements, techno-economic considerations, and problem formulation.
The value, effectiveness, and efficiency of the proposed framework are evaluated by formulating and solving three different case studies in Sec.~\ref{sec:section3}, each deliberately selected to highlight specific aspects of HES design.
Finally, Sec.~\ref{sec:conclusion} offers some remarks regarding conclusions, limitations, and future directions.
These sections are complemented by additional explanations in the Appendix (Sec.~\ref{sec:Appendix}).
These include a nomenclature, node definitions and mathematical descriptions (associated with Fig.~\ref{fig:IES}), optimization model, and lexical interpretations for problem elements.
We recommend referring to these resources for further clarification.%

\xsection{Problem Description}\label{sec:section2}

In this section, we describe the HES architecture, problem elements, techno-economic considerations, and the capacity and dispatch optimization problem. 
The framework develops each subsystems' fundamental dynamics based on basic energy balance and operating states along constraints representing basic limitations to develop a linear program that can be efficiently solved and is enabled by prior work in Refs.~\cite{Vercellino2022}.

\subsection{HES Architecture}\label{subsec:Architecture}

Figure~\ref{fig:IES} describes the general architecture of a hybrid energy system (HES) considered in this framework.
An HES is characterized by various generators (e.g.,~NGCC power plant, wind farm, or NPP) and storage units (e.g.,~TES, BESS, or hydrogen storage).
Additional functions (e.g.,~CCS, district heating/cooling, etc.) may also be present in the system and are characterized by the addition of their associated costs and energy requirements.

The generator, which is described by the subscript $\parm_{G}$, is the unit responsible for producing electricity from an energy vector (e.g.,~natural gas) by often converting it first to a primary energy domain (e.g.,~thermal energy), and ultimately converting it to electricity with efficiency of $\eta_{G}$:
\begin{align}
    \label{Eq:Gen_eff}
    \text{generator's primary energy domain} \xrightarrow{\eta_{G}}\text{electricity}
\end{align}

\noindent
Based on the desired configuration, the generator may have access to three different types of storage systems, distinguished through the subscript $\parm_{S}$.
These storage systems are defined based on their key energy domain.
The first storage system uses the generator's primary domain, $\parm_{P}$, to store energy.
As an example, in an NGCC power plant, the primary energy domain is thermal energy.
Thus, the primary storage system for this generator is a TES unit.
In this work, the secondary energy domain is always electrical, $\parm_{E}$.
Therefore, the second storage type directly stores energy as electricity (e.g., through a BESS).
A third storage facility, $\parm_{T}$, is included to enable the usage of electricity to create a new product, commodity, service, or energy storage in a distinct medium, such as hydrogen.
This architecture is primarily motivated by a range of potential energy storage conceptual areas that map to both new and upcoming technologies \cite{rahman2020assessment}.
Therefore, this HES architecture supports the inclusion of more than a single solution of interest.

The storage system is characterized by charge $\protect\overrightharpoon{\parm}$ and discharge $\overleftharpoon{\parm}$ signals (noting the arrow directions).
The charge signal can be affected by losses during transmission.
Therefore, its efficiency is denoted by $\overrightharpoonsub{\eta}{\parm}$.
The output efficiency of the storage is described as $\overleftharpoonsub{\eta}{\parm}$:
\begin{subequations}
\label{Eq:Str_eff}
\begin{align}
    \text{charging signal} &\xrightarrow{\overrightharpoonsub{\eta}{\parm}}\text{storage} \\
    \text{storage}  &\xrightarrow{\overleftharpoonsub{\eta}{\parm}} \text{discharge signal}    
\end{align}
\end{subequations}

This systematic separation of efficiencies allows the framework's users to consider additional factors, such as the distance between the facilities, that can affect the efficiency of energy transmission.
An additional revenue-driven control signal, represented by $\parm_{R}$, is responsible for deciding the percentage of the discharge signal that is immediately turned into revenue through direct sales.
For example, the revenue signal will decide how much of the discharged hydrogen is sold at current hydrogen prices.
The remaining product will be combusted and sold as electricity.

The inclusion of various functions may result in the presence of additional load requirements. 
For example, in addition to auxiliary electrical loads, the operation of a CCS unit, which is assumed to be in operation whenever the generator is on, requires additional thermal and electrical loads \cite{Vercellino2022}.
These loads are a function of the generator's current power level, presented as a certain percentage of the power plant's power level.
Similarly, a high-temperature steam electrolysis (HTSE) process in a hydrogen plant (tertiary storage) requires both thermal and electrical energy \cite{Mingyi2008}. However, these load requirements are active only when a decision is made to generate hydrogen from the excess electricity.
Therefore, these loads are a function of the tertiary charging signal, presented as a certain percentage of its current value.

The top part of Fig.~\ref{fig:IES} highlights some considerations regarding the electric energy technologies, including fuel cost, carbon tax, and dispatchable versus non-dispatchable types of resources.  
The bottom part presents a case for a collection of homogeneous generators (e.g.,~a wind farm) in the presence of multiple functions and storage types. 
The nodes in this figure, which are described by $n_{1}$, $\dots$, $n_{9}$ are used to formulate some of the necessary constraints within the optimization problem and are mathematically described in Appendix~\ref{sec:Appendix_nodes}.

\subsection{Problem Elements}\label{subsec:ProblemElements}

This section introduces some problem elements based on the comprehensive case in which a collection of homogeneous generators has potential access to all $3$ storage types in the presence of both primary and electrical loads.
Adding the capability of simultaneously working with multiple non-homogeneous generators will be a future step of this work.
Similarly, in the future, we plan to enable a more advanced integration of storage units in the toolbox, such that multiple storage topologies (e.g.,~two battery storage and one hydrogen storage) can be simultaneously integrated with \ToolName.
A lexical interpretation of the problem elements is included in Tab.~\ref{Tab:lexicalInterp} in the Appendix (Sec.~\ref{sec:Appendix_lex}).

\subsubsection*{Plant Variables:}
The capacity of each storage type is a sizing decision that will be determined by the optimization algorithm and constitutes the plant optimization variables $\bm{\Sigma}$:
\begin{align}
    \label{eq:plantvar}
     \bm{\Sigma} = [\Sigma_{P}, \Sigma_{E}, \Sigma_{T}]^{T}
\end{align}

\noindent
Note that the vector of plant optimization variables reduces in size if the study does not include all three storage types.

\subsubsection*{Control Variables:}~Every energy storage system entails $3$ control variables, one for charging the storage ($\overrightharpoonsub{u}{\parm}$), one for discharging the storage $\overleftharpoonsub{u}{\parm}$, and one for determining the fraction of discharge $u_{\parm R}$ that is directly used to generate revenue without any intermediate steps. 
In addition, the operator can request a specific power from the generator through a control command, described as $u_{G}(t)$.
The vector of control variables can then be defined as:
\begin{equation}
    \begin{aligned}
        \label{eq:Controlvar}
        \bm{u}(t) & = [u_{G}(t), \bm{u}_{S}(t)]^{T}\\
        &  = [u_{G}(t), \bm{u}_{SP}(t), \bm{u}_{SE}(t), \bm{u}_{ST}(t)]^{T}
    \end{aligned}
\end{equation}

\noindent
Here, every storage control vector $\bm{u}_{S\parm}(t)$ consists of $3$ variables $\bm{u}_{S\parm}(t) = [\,\overrightharpoonsub{u}{\parm}(t), \overleftharpoonsub{u}{\parm}(t), u_{\parm R}(t)]^{T}$.
Similar to the previous case, the size of the control vector will be reduced if only some of the storage types are included in the study.

\subsubsection*{State Variables:}~There is one state variable associated with the generator which describes the power level of the generator using a ramp rate of $\tau$:
\begin{equation}
    \label{eq:generatordynamics}
    \dot{x}_{G}(t) = \frac{1}{\tau}(-x_{G}(t)+u_{G}(t))
\end{equation}

\noindent
Each storage system is characterized by a state variable that describes the current amount of stored energy in that system.
For the most comprehensive case with $3$ different storage facilities, the storage dynamics are described by:
\vspace{2\baselineskip}
\begin{align}
 \label{eq:Storagedynamics}
 \dot{\bm{x}}_{S}(t) = \begin{bNiceMatrix}
        \dot{x}_{P}(t)\\
        \dot{x}_{E}(t) \\
    \dot{x}_{T}(t) \\
    \end{bNiceMatrix}
    = \begin{bNiceMatrix}
       \overrightharpoonsub{\eta}{P} \overrightharpoonsub{u}{P}(t)\\
        \overrightharpoonsub{\eta}{E} \overrightharpoonsub{u}{E}(t) \\
        \alpha_{ET} \overrightharpoonsub{\eta}{T} \overrightharpoonsub{u}{T}(t) \\
        \CodeAfter
        \OverBrace[shorten,yshift=3pt]{1-1}{3-1}{\text{Charge}}
    \end{bNiceMatrix}
    - \begin{bNiceMatrix} \overleftharpoonsub{u}{P}(t)\\
       \overleftharpoonsub{u}{E}(t) \\
       \overleftharpoonsub{u}{T}(t)
      \\
        \CodeAfter
        \OverBrace[shorten,yshift=3pt]{1-1}{3-1}{\text{Discharge}}
       \end{bNiceMatrix}
\end{align}

\noindent
where $\alpha_{ET}$ is the conversion rate between electricity and the tertiary commodity. 
Describing storage states in vector form and augmenting them with the generator state, the dynamics of the problem are described by: 
\begin{align}
    \label{eqn:Dynamics1}
    \dot{\bm{x}}(t) & =  \bm{A}\bm{x}(t) + \bm{B}\bm{u}(t) \\
    \begin{bNiceMatrix}
        \dot{x}_{G}(t) \\
        \dot{\bm{x}}_{S}(t) 
        \end{bNiceMatrix} & =  \begin{bNiceMatrix}
        -1/\tau & \bm{0}\\
        \bm{0} & \bm{0} 
    \end{bNiceMatrix} \begin{bNiceMatrix}
        x_{G}(t)\\
        \bm{x}_{S}(t)
    \end{bNiceMatrix}
    + \begin{bNiceMatrix}
        1/\tau & \bm{0}\\
        \bm{0} & \bm{b}_{S}
    \end{bNiceMatrix} \begin{bNiceMatrix}
        u_{G}(t)\\
        \bm{u}_{S}(t)
    \end{bNiceMatrix} \notag
\end{align}

\noindent
where $\bm{b}_{S}$ is the appropriately-sized matrix:
\begin{align}
\label{eqn:Dynamicsbs}
\bm{b}_{S} = 
\begin{bNiceMatrix}[columns-width = 5mm]
        {\overrightharpoonsub{\eta}{P}} & -1 & 0 & 0 & 0 & 0 & 0 & 0 & 0 \\ 
        0 & 0 & 0 & \overrightharpoonsub{\eta}{E} & -1 & 0 & 0 & 0 & 0\\ 
        0 & 0 & 0 & 0 & 0 & 0 & \alpha_{ET}\overrightharpoonsub{\eta}{T} & -1 & 0
\end{bNiceMatrix}
\end{align}

\subsubsection*{Constraints:}\label{subsec:Constraints}~This section presents all of the time-independent and time-dependent constraints in the dynamic optimization problem.

The storage capacities are non-negative.
Therefore, the following constraint is imposed on plant variables:
\begin{align}
    \label{eq:plantcons1}
    \bm{0} \leq \bm{\Sigma} 
\end{align}

The requested power from the generator is non-negative and less than or equal to the net nominal capacity of the generator.
Similarly, the charge and discharge signals are non-negative and never greater than the maximum energy transfer rate into and out of the storage system, respectively. 
These maximum and minimum energy transfer rates are currently input parameters but will be added to the set of potential plant optimization variables in the future, similar to Ref.~\cite{Vercellino2022}.
These constraints are succinctly described in vector form as:

\begin{align} \label{eq:controlcons1}
    \bm{0} \leq \bm{u}(t) \leq \bm{u}_{\text{max}}(t) 
\end{align} 
\noindent Specifically, the revenue-generating fraction of the control signal in Eq.~(\ref{eq:controlcons1}) is non-negative and equal to or smaller than the discharge signal: 

\begin{subequations}
\label{eq:discharge}
\begin{align}
    u_{PR}(t) & \leq \overleftharpoonsub{u}{P}(t) \label{Eq:P_rev}\\
    u_{ER}(t) & \leq \overleftharpoonsub{u}{E}(t) \label{Eq:E_rev}\\
    u_{TR}(t) & \leq \overleftharpoonsub{u}{T}(t) \label{Eq:T_rev}
\end{align}
\end{subequations}

The generator's power level must remain non-negative and never exceed its nominal capacity.
Requesting a specific (admissible) power output from the generator is reasonable for technologies such as nuclear or NGCC under simplifying assumptions like no temperature dependence on its operation or maintenance schedules.
For intermittent technologies, such as wind and solar, they are at the mercy of the availability of (renewable) resources. 
In other words, the electricity produced by such technologies is not dispatchable due to the resource's inherent intermittency.
Thus, it is necessary to ensure that the generator state $x_{G}(t)$ is bounded by an input signal that represents the level of resource availability.
These result in the following constraint on the generator's state:
\begin{align}
    \label{eq:controlcons2}
   \bm{x}_{G,\text{min}}(t) \leq \bm{x}_{G}(t) \leq \bm{x}_{G,\text{max}}(t)
\end{align}

\noindent
where $\bm{x}_{G}(t)$ refers to the generator's state, and $\bm{x}_{G,\text{max}}(t)$ is an upper bound that is established based on nominal capacity, or the availability of renewable resources. 
Further details regarding the construction of $\bm{x}_{G,\text{max}}(t)$ for a wind farm are included in Sec.~\ref{subsec:CaseStudy2}.

The amount of stored energy must be non-negative and less than or equal to the capacity of the storage system: 
\begin{equation}
    \label{eq:storagecons}
    \bm{0} \leq \bm{x}_{S}(t) \leq \bm{\Sigma}
\end{equation}

\noindent 
which represents a key coupling between select states and the plant parameters.

The initial states are prescribed for all the state variables. It can also be assumed that at the final time $t_{f}$, the storage system has the same amount of stored energy as at $t_{0}$:
\begin{subequations}
\label{Eq:initfinal}
\begin{align}
    \label{eq:InitFinal}
    \bm{x}(t_{0}) &= \bm{x}_{0}\\
    \bm{x}_{S}(t_{f}) &= \bm{x}_{S}(t_{0})
\end{align}
\end{subequations}

\noindent
where the latter equation is optional as multiple shorter, sequential time horizons do not necessitate this assumption~\cite{Vercellino2022}.

In addition to the upper bound imposed by the maximum energy transfer rate for charging the storage system in Eq.~(\ref{eq:controlcons1}), it is necessary to ensure that the charging signal is smaller than or equal to the available power in the generator.
This is described with the help of nodes that are placed in Fig.~\ref{fig:IES}:
\begin{subequations}
\label{Eq:n146}
\begin{align}
    \overrightharpoonsub{u}{P}(t) & \leq n_{1}(t) \label{Eq:P_n1}\\
    \overrightharpoonsub{u}{E}(t) & \leq n_{5}(t)  \label{Eq:P_n5}\\
    \overrightharpoonsub{u}{T}(t) & \leq n_{7}(t)  \label{Eq:P_n6}
\end{align}
\end{subequations}

\noindent
where the mathematical expressions associated with all of the nodes $n_{1}$, $\dots$, $n_{9}$ are provided in Appendix~\ref{sec:Appendix_nodes}. 

In addition, the generator's load-satisfying signals $L_{GP}$, $L_{GE}$, and $L_{PT}$ are non-negative.
This condition is to ensure that power does not flow from the storage system into the generator.
These signals are also upper-bounded by the available power in the generator. 
These constraints are formulated as:
\begin{subequations}
\label{eq:LGPLGE}
\begin{align}
     \bm{0} &\leq L_{GP}(t) \leq n_{2}(t) \label{Eq:LGP}\\
     \bm{0} & \leq L_{GPT}(t) \leq n_{3}(t) \label{Eq:LPT}\\
     \bm{0} &\leq L_{GE}(t) \leq n_{5}(t) \label{Eq:LGE} 
\end{align}
\end{subequations}

\subsubsection*{Objective Function:}\label{subsec:Objective}~The net present value (NPV) objective function, which enables the assessment of the economic viability of a given technology, is used in this framework.
NPV is calculated as:
\begin{align}
    \label{eq:NPV}
    \text{maximize} \quad \text{NPV} & = -C_{\text{cap}}{(\bm{\Sigma})} + \int_{t_0}^{t_f} \frac{v_{\text{profit}}({\bm{u}, \bm{x}, \bm{\Sigma}}, t)}{D(t)} dt
\end{align}

\noindent 
where $C_{\text{cap}}$ are the capital expenses, $v_{\text{profit}}(t)$ is calculated as a function of expenses and revenues, and $D(t)$ is the discounting function (money is `worth' more now than in the future).
An annualized discounting function is considered here as:
\begin{align}
D(t) = (1 + r)^{\text{year}(t)}
\end{align}

\noindent
where $r$ is the discount rate and $\text{year}(t)$ is the integer number of years that have passed since $t_0$.
These intermediate quantities are now discussed in detail.

\subsection{Techno-Economic Considerations}\label{subsec:Techno-Economics}

In order to construct the NPV objective function in Eq.~(\ref{eq:NPV}), we first consider the sources of costs and revenues.
These cost parameters are a user-defined input to the \ToolName~framework.

\subsubsection{Expenses}\label{subsubsec:expenses}~All of the costs included in the techno-economic analysis (including capital costs, fixed and variable operation and maintenance, fuel costs, and carbon costs) are described in this section. 

\paragraph{Capital Costs:}\label{subsubsec:Cap}~In this article, we assume that the capital cost, $C_{\text{cap}}$ consists of overnight capital costs $C_{\text{occ}}$, and costs over the period of construction $C_{cp}$:
\begin{align}
    \label{eq:cap}
    C_{\text{cap}} = C_{\text{occ}} + C_{cp}
\end{align}

\noindent
where $C_{\text{occ}}$ assumes that all of the construction occurs overnight.
Therefore, this term excludes changes in the prices of goods and financial costs (such as the loan, inflation, discount rate, etc.).
This allows potential investigations into the impact of construction periods, rates of inflation, etc., in the analysis \cite{Joskow2009}.
$C_{\text{occ}}$ consists of direct construction costs, indirect construction costs, contingencies, and the owner's cost, explained in detail in Ref.~\cite{Mantripragada2018, international2013iaea}.

$C_{cp}$ includes all the costs that are incurred over the construction period, such as escalation, loan, inflation, etc.
Therefore, this term is sensitive to the choice of financial parameters such as discount rate, debt-equity ratio, interest rate, interest during construction (IDC), etc.
In this study, we simplify this term, similar to the methodology presented in Ref.~\cite{Wealer2021}, to account for the costs over the period of construction through a simple model characterizing IDC:
\begin{align}
    \label{eq:cap_idc0}
    C_{\text{cap}} = C_{\text{occ}}(1+C_{\text{idc}})
\end{align}

\noindent
where $C_{\text{idc}}$ is calculated as a function of the construction time $T_{\text{con}}$, and the cost of capital rate $r$, estimated as:
\begin{equation} \label{eq:cap_idc}
    C_{\text{idc}} = \frac{r}{2}T_{\text{con}} + \frac{r^2}{6}T_{\text{con}}^{2}
\end{equation}

\noindent
Inclusion of more advanced financial parameters \cite{Vercellino2022}, such as loan, depreciation, etc., is a future work item for this framework.
For the entire system, the capital cost is expressed as: 
\begin{align}
    \label{eq:cap_final}
    C_{\text{cap}} = [C_{\text{occ}_{G}} + C_{\text{occ}_{P}}\Sigma_{P} +  C_{\text{occ}_{E}}\Sigma_{E} + C_{\text{occ}_{T}}\Sigma_{T}](1+C_{\text{idc}})
\end{align}

\paragraph{Operation and Maintenance (O\&M):}\label{subsubsec:OM}~The main elements included in O\&M costs are associated with fixed and variable O\&M costs.
Fixed O\&M costs, expressed as $C_{\text{fom}}$, include regular system maintenance, decommissioning, component replacement, etc.
These costs only depend on the duration of the operation. 
For storage systems, these costs scale with storage size (e.g.,~more frequent and lengthier maintenance for larger capacities).
Fixed O\&M costs $C_{\text{fom}}$ are then calculated as:
\begin{align}
    \label{eqn:fomcost}
    C_{\text{fom}}(t) & = C_{\text{fom}_{G}} + C_{\text{fom}_{P}}\Sigma_{P} + C_{\text{fom}_{E}}\Sigma_{E} + C_{\text{fom}_{T}}\Sigma_{T}
\end{align}

Variable O\&M costs, expressed as $C_{vom}$, reflect the non-fuel portion of the costs that vary by the amount of energy generated or supplied (such as water, waste disposal, lubricants, chemicals, and other consumable materials).
Therefore, these costs depend on the power level of the unit, expressed as:
\begin{subequations}
\label{eqn:vomcost}
\begin{align}
    C_{\text{vom}}(t)  = C_{\text{vom}_{G}}\bm{x}_{G}(t) &+ C_{\text{vom}_{P}}(\,\overrightharpoonsub{\eta}{P}\overrightharpoonsub{u}{p}(t) + \overleftharpoonsub{u}{p}(t)\,) \\
    & + C_{\text{vom}_{E}}(\,\overrightharpoonsub{\eta}{E}\overrightharpoonsub{u}{E}(t) + \overleftharpoonsub{u}{E}(t)\,) \\
    & + C_{\text{vom}_{T}}(\,\alpha_{ET}\overrightharpoonsub{\eta}{T}\overrightharpoonsub{u}{T}(t) + \overleftharpoonsub{u}{T}(t)\,)
\end{align}
\end{subequations}

\paragraph{Fuel Costs:}\label{subsubsec:fuel}~Generally, fuel cycle cost, $C_{\text{fuel}}(t)$ includes both front-end and back-end costs, such as supply, conversion, enrichment, fabrication, transportation, and waste disposal.
For fossil fuels, the cost of fuel only entails the front-end costs, as received from the market.
However, for nuclear power plants, the back-end costs are also included as a percentage of the front-end costs:
\begin{align}
    \label{eq:fuelcost}
    E_{\text{fuel}}(t) = \rho_{\text{fuel}} C_{\text{fuel}}(t)x_{G}(t) + B_{\text{fuel}}(t)
\end{align}

\noindent
where $\rho_{\text{fuel}}$ is the conversion factor between the power output of the generator and fuel consumed,
$C_{\text{fuel}}(t)$ is the instantaneous fuel price, and $B_{\text{fuel}}(t)$ is the back-end cost (again, mainly necessary for waste management and disposal in nuclear power plants).

\paragraph{Carbon Cost:}\label{subsubsec:carbon}~With the goal of making the societal cost of carbon emissions visible, carbon cost, expressed as CO\textsubscript{2}, assumes a carbon tax rate to incentivize clean electricity generation.
The cost of fuel is therefore estimated as:
\begin{align}
    \label{eq:carboncost}
    E_{\text{CO\textsubscript{2}}}(t) = C_{\text{CO\textsubscript{2}}} \alpha_{\text{CO\textsubscript{2}}}\rho_{fuel}x_{G}(t)
\end{align}

\noindent
where $C_{\text{CO\textsubscript{2}}}$ is the carbon tax, and $\alpha_{\text{CO\textsubscript{2}}}$ is the amount of Co$_2$ produced per unit of fuel.

\subsubsection{Revenue}\label{subsubsec:revenues}:~Here, the operator can earn revenue by either directly selling electricity to the grid (without using storage), or alternatively, using storage to sell primary energy, electricity, or a tertiary commodity.
The total revenue is calculated for all energy domains as a function of their associated prices.
The price arbitrage between primary, electricity, and tertiary domains determines the optimal flow of energy that maximizes the objective function.
As an example, revenue earned by selling stored thermal energy to chemical plants (first term in Eq.~(\ref{eq:revenue})) is calculated as a function of thermal energy prices $C_{P}(t)$, discharge efficiency $\overleftharpoonsub{\eta}{P}$, and the revenue control signal $u_{PR}(t)$.
However, the revenue control signal may only become active when the electricity prices are low and thermal energy prices are high. 
Note that for certain commodity types, such as hydrogen, it is also possible to generate electricity, which will then be sold to the grid.
Mathematically, the revenue  can be described as: 
\begin{align}
    \label{eq:revenue}
   R   = &~R_{P}(t) + R_{E}(t) + R_{T}(t) \\
       = &~C_{P}(t)\overleftharpoonsub{\eta}{P}u_{PR}(t) + C_{E}(t)\eta_{G}x_{G}(t)- C_{E}(t)\eta_{G}\overrightharpoonsub{u}{P}(t) \notag \\
       &~- C_{E}(t)\eta_{G}L_{P}x_{G}(t) + C_{E}(t)\eta_{G}\overleftharpoonsub{\eta}{P}\overleftharpoonsub{u}{P}(t) \notag \\
       &~-C_{E}(t)\eta_{G}\overleftharpoonsub{\eta}{P}u_{PR}(t) - C_{E}\eta_{G}L_{PT}\overrightharpoonsub{u}{T} -   C_{E}(t)\overrightharpoonsub{u}{E}(t) \notag \\
       &~-C_{E}(t)L_{E}x_{G}(t) + C_{E}(t)\overleftharpoonsub{\eta}{E}\overleftharpoonsub{u}{E}(t) - C_{E}(t)\overrightharpoonsub{u}{T}(t) \notag \\
       &~+\alpha_{\text{TE}}C_{E}(t)\overleftharpoonsub{\eta}{T}\overleftharpoonsub{u}{T}(t) -\alpha_{\text{TE}}C_{E}(t)\overleftharpoonsub{\eta}{T}u_{TR}(t) \notag \\
       &~+ C_{T}(t)\overleftharpoonsub{\eta}{T}u_{TR}(t) \notag
\end{align}

\noindent
where $C_{P}(t)$ is the price of the primary energy, $C_{E}(t)$ is the electricity price, and $C_{T}(t)$ is the price of tertiary commodity/product.
In this equation, the first term, $C_{P}(t)\overleftharpoonsub{\eta}{P}u_{PR}(t)$, is the revenue earned by selling stored energy from primary storage.
The last term, $C_{T}(t)\overleftharpoonsub{\eta}{T} u_{TR}(t)$, is the revenue earned by selling stored tertiary energy or product.
All the terms in between are related to revenue gained from electricity sales, taking into account that a portion of the electricity generated by the generator might be used for various purposes (e.g.,~loads for function requirements or tertiary conversion).

\subsection{Problem Formulation}\label{subsec:OptimizationFormulation}
With all of the problem elements defined, the economic feasibility of a candidate HES can be assessed through the optimization of capacity and dispatch within an all-at-once problem formulation: 
\begin{subequations}
\label{Eqn:Mainoptform}
\begin{align}
\textrm{changing:} \quad & \bm{u}(t), \bm{x}(t), \bm{\Sigma} \\
\textrm{maximize:} \quad & \text{NPV}(t,\bm{u},\bm{x}, \bm{\Sigma},\bm{d}) \label{Eqn:Objoptform}\\
\textrm{subject to:} 
\quad & \bm{g}(t,\bm{u},\bm{x}, {\bm{\Sigma}},\bm{d})\leq \bm{0}\label{Eqn:Ineqoptform}\\
\quad & \dot{\bm{x}} - \bm{f}(t,\bm{u},\bm{x}, {\bm{\Sigma}},\bm{x}_{0},\bm{d}) = \bm{0} \label{Eqn:Dynoptform}\\ 
& \bm{x}(t_{0}) = \bm{x}_{0}\\
& \bm{x}(t_{f}) = \bm{x}(t_{0}) \ \text{(optional)}\\
\textrm{where:} \quad
& \bm{u} = \bm{u}(t),\ \bm{x} = \bm{x}(t) \\
& t \in[t_{0},t_{f}],\ t_{f} = Nt_{p} \label{eq:PC} 
\end{align}
\end{subequations}
 
\noindent
where $\bm{d}$ is the vector of problem parameters, $\bm{g}(\cdot)$ is the vector of inequality constraints, associated with
 Eqs.~(\ref{eq:plantcons1})--(\ref{Eq:LGE}) and Eq.~(\ref{Eqn:Dynoptform}) refers to systems dynamics described in Eqs.~(\ref{eq:generatordynamics}) and (\ref{eq:Storagedynamics}).
Periodic conditions are defined in Eq.~(\ref{eq:PC}) using a base period $t_{p}$, and the number of repetitions, $N$.
Periodic conditions can be adjusted by the user based on the availability of price signals, time horizon of the problem of interest, and other factors. 
For further clarification, this formulation is accompanied by Tab.~\ref{Tab:lexicalInterp}, which provides a lexical interpretation for some of the problem elements.
In addition, an optimization model presented in Fig.~\ref{fig:OptModel}, is created in Sec.~\ref{sec:Appendix} to illustrate the relationship between various problem elements better.    

As it is well established in mathematical optimization literature, the choice of units can negatively affect the problem's scaling and decrease its effectiveness \cite{Martins2021}.
Appropriately adjusting the order of magnitude of problem elements can result in a more computationally favorable problem, preventing unstable and inefficient algorithmic calculations \cite{Herber2017}.  
As an example, solving the case studies in this article without appropriate scaling using a $10^{-6}$ solver tolerance can take up as much as a day of computational time.
This is because the optimization algorithm will spend a tremendous amount of time optimizing for $10^{-6}$th of a dollar value with respect to the gradient of the objective function.
From a broader perspective, it is clear that $10^{-6}$th of a dollar value is insignificant over the project's lifetime, which has an NPV value in the order of billions of dollars.  
Using a scaling factor of $10^9$ for the objective function and appropriately-selected scaling factors for other problem elements, the problem can be solved in less than $1800$ seconds, signifying a dramatic increase in computational efficiency. 
A more comprehensive discussion on scaling in dynamic optimization problems is discussed in Ref.~\cite{Herber2017}.

\section{Case Studies}\label{sec:section3}

\begin{figure*}[t]
\centering
    \captionsetup[subfigure]{justification=centering}
    \centering
    \begin{subfigure}{0.333\textwidth}
    \centering
    \includegraphics[scale=0.45]{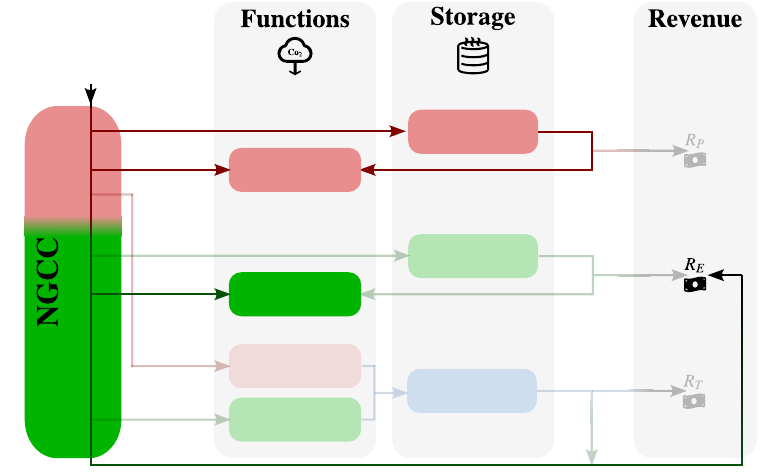}
    \caption{Case Study \rom{1}.}
    \label{fig:IES_CaseI}
    \end{subfigure}%
    \begin{subfigure}{0.333\textwidth}
    \centering
    \includegraphics[scale=0.45]{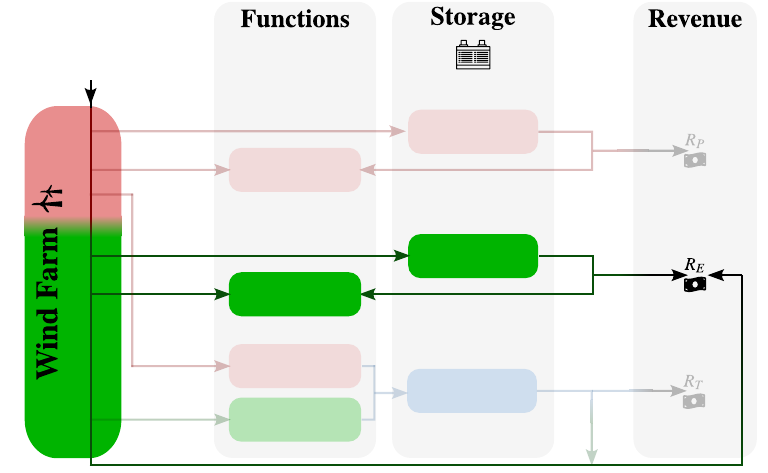}
    \caption{Case Study \rom{2}.}
    \label{fig:IES_CaseII}
    \end{subfigure}%
    \begin{subfigure}{0.333\textwidth}
    \centering
    \includegraphics[scale=0.45]{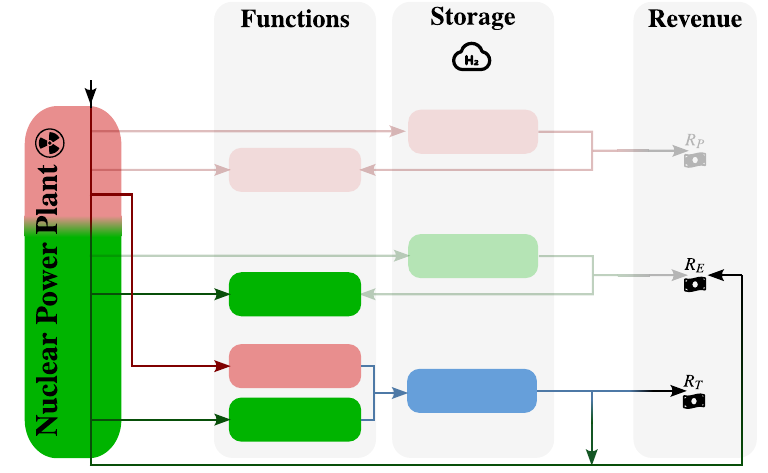}
    \caption{Case Study \rom{3}.}
    \label{fig:IES_CaseIII}
    \end{subfigure}%
    \captionsetup[figure]{justification=centering}
    \caption{HES candidate for Case Study~\rom{1}: a natural gas combined cycle power plant with thermal storage and a carbon capture and storage system, Case Study~\rom{2}: a wind farm with a battery energy storage system, and Case Study~\rom{3}: a nuclear power plant with a hydrogen production (through high-temperature steam electrolysis) and storage facility.}
    \label{fig:IES_Cases}
\end{figure*}

This section shows several case studies to demonstrate the capabilities of the proposed framework. 
These studies are selected to highlight, to the extent possible, different modes of operation using different technologies and storage types.
Shared parameters among these technologies are described in Table~\ref{tab:par1}.
A brief summary of each study is discussed next:
\begin{enumerate}[topsep=0pt,itemsep=-1ex,partopsep=1ex,parsep=1ex,label=$\bullet$]
    
\item \textit{Case Study~\rom{1}}: NGCC power plant with CCS and TES. 
This study examines an architecture with \textit{primary} storage and CCS function, signifying the effectiveness of the proposed framework in alleviating negative implications associated with the parasitic load required to operate the CCS system \cite{Limb2022,  Vercellino2022, markey2024economic}.
This case study also highlights optional load-following requirements.%

\item \textit{Case Study~\rom{2}}: Wind farm with a BESS. 
As the second-most popular proposed hybrid, this wind-based system focuses on variable renewable energy generation \cite{Schleifer2023} combined with a \textit{secondary} storage unit.
In addition to common use in HES \cite{yi2022energy, gwabavu2021dynamic, naemi2022optimisation}, this study highlights how non-dispatchable loads can be integrated into the framework and illustrates assessing numerous scenarios based on the location (and associated wind profiles) to guide the decision-making process.

\item \textit{Case Study~\rom{3}}: Nuclear power plant with hydrogen generation.
This study examines a common HES application that employs a \textit{tertiary} storage system to generate non-energy products \cite{Frick2019, Zhang2022}.
By focusing on non-energy products, this study provides insights on how HESs impact both energy and non-energy markets, such as with time-based market constraints \cite{Frick2019}.
Sensitivity analysis is also shown for storage overnight construction costs and hydrogen prices.

\end{enumerate}

\ToolName~uses an hourly time mesh within DTQP by default, although other time intervals (e.g., decisions being made every minute) can be used as desired.  
The control decisions over these intervals are piecewise constant, making the zero-order hold method suitable, as it produces no discretization error for the state dynamics. 
The resulting CCD optimization problem, with the current assumptions outlined in the previous section, is a linear optimization problem.
Due to the problem's convexity property, it can be efficiently solved for the global optimal solution using \texttt{MATLAB}'s linprog optimization solver (although other solvers could be used given the problem matrices from DTQP).
A solver tolerance of $10^{-9}$ was used in linprog, with the default dual-simplex-highs algorithm to solve a dynamic optimization problem using direct transcription with an equidistant mesh and a composite Euler forward quadrature method.

Problem setup and solving times presented are associated with a single desktop workstation with an AMD Ryzen 9 3900X 12-core processor at 3.79 \unit{GHz}, 32 \unit{GB} of RAM, 64-bit Windows 10 Enterprise LTSC version 1809, and \texttt{Matlab} R2024a.

\begin{table*}[t]
\caption{Cost parameters for generator and storage technologies used in Case Study \rom{1}, \rom{2}, and \rom{3}.}
        \label{tab:par1}
    \renewcommand{\arraystretch}{1.1}
    \centering
    \begin{tabular}{r| r r r r| r r r r r r }
        \hline  \hline
         \multirow{2}{*}{\text{Parameters}} & \multicolumn{4}{c}{\text{Generator}} & \multicolumn{5}{c}{\text{Storage}}\\
         \cline{2-5} \cline{6-10}
         & \text{CC} \cite{energy2020capital} & \text{Wind} \cite{energy2020capital} & \text{Nuclear}~\cite{energy2020capital} &  \text{Unit} & \text{TES} \cite{Vercellino2022} & \text{BESS} \cite{energy2020capital} & \text{Unit} & \text{Hydrogen}~\cite{Frick2019} & \text{Unit} \\
         \hline
         \text{Nom. cap.} & $1083$ & $200$ & $2156$ &  \unit{MW} & - & $50$ & \unit{MW} & $640$ & \unit{tpd}  \\
         $C_{\text{occ}}$ &  $0$  & $1265000$ & $6041000$ & \unit{\$/MW} & $1048947$ & $173500$ & \unit{\$/MWh} & $600.074$ &  \unit{\$/kg} \\
         $C_{\text{fom}}$ & $12200$ & $26340$ & $121640$ & \unit{\$/MW-yr} & $4.7897$ & $0.7178$  &  ~\unit{\$/MWh-h} & $0$ & \unit{\$/kg-yr} \\
         $C_{\text{vom}}$ & $1.87$ & $0$ & $2.37$ & \unit{\$/MWh} & $0.75$ & $0$ & \unit{\$/MWh} & $0.2884$ & \unit{\$/kg} \\
         \hline
    \end{tabular}
\end{table*}

\subsection{Case Study \rom{1}: Combined Cycle with Thermal Storage and Carbon Capture}
\label{subsection:CS1}

To meet potential environmental requirements, fossil fuel-based generators, such as NGCC power plants, are considering a reduction in their carbon emissions through CCS functions.
However, the inclusion of a CCS reduces the net plant efficiency and power output while increasing the cost of electricity \cite{Rubin2012}. 
A potential mitigating solution is to integrate the system with a thermal energy storage (TES), as discussed in detail in Refs.~\cite{Vercellino2022, Limb2022}.

This study assumes that the NGCC power plant, which is already built and equipped with a CCS unit (therefore, there is no overnight construction cost), is to be retrofitted to include a hot TES system.
The CCS operates only when the plant is running.
The CCS function is characterized by a thermal and electrical load, which are required for the operation of the CCS unit and are defined as a fixed percentage of the generator's power level.
Therefore, whenever the generator is on, these electrical and thermal loads appear in the problem and must be satisfied. 
Including a hot thermal storage unit in the architecture enables operators to practically remove some of the parasitic thermal load from the power plant when the electricity prices are high, maximizing revenue.
The considered system is illustrated in Fig.~\ref{fig:IES_CaseI}, and specific, technology-dependent parameters are shown in Tab.~\ref{tab:CaseIPar}.

\begin{table}[ht]
    \centering
    \caption{Parameters associated with Case Study \rom{1} for combined cycle generator with CCS, and a thermal storage system largely based on Ref.~\cite{Vercellino2022}.}
        \label{tab:CaseIPar}
    \renewcommand{\arraystretch}{1.1}
    \begin{tabular}{r| r r r| r r}
        \hline  \hline
         \textbf{Field} & {\textbf{Value}} & \textbf{Unit} & \textbf{Field} & {\textbf{Value}} & \textbf{Unit}\\
         \hline
          $\rho_{\text{fuel}}$ & $146.952$ &\unit{kg/h.MW} & $\alpha_{\text{CO\textsubscript{2}}}$ & $0.0029$ & \unit{ton/kg } \\
          $\tau$ & $0.1389$ & \unit{h} & $\eta_{G}$ & $1$ & - \\
          $u_{G,\text{min}}$ & $0$ & \unit{MW} & $u_{G,\text{max}}$ & $1083$ & \unit{MW}\\
          $x_{G,\text{min}}$ & $0$ & \unit{MW}& $x_{G,\text{max}}$ & $1083$ & \unit{MW}\\
          $x_{S}(t_{0})$ & $25$ & \unit{MWh}& $x_{S}(t_{f})$ & $25$ & \unit{MWh}\\
          $\overrightharpoonsub{u}{\text{max}}$ & $200$ & \unit{MW} & $\overleftharpoonsub{u}{\text{max}}$ & $200$ & \unit{MW}\\
          $L_{P}$ & $0.1 x_{G}$ & \unit{MW} & $L_{E}$ & $0.2 x_{G}$ & \unit{MW}\\
          $T_{\text{con}}$ & $3$ & \unit{years} & $r$ & $0.075$ & -\\
         \hline
    \end{tabular}
\end{table}

\begin{figure*}[t]
    \captionsetup[subfigure]{justification=centering}
    \centering
    \begin{subfigure}{0.25\textwidth}
    \centering
    \includegraphics[scale=0.49]{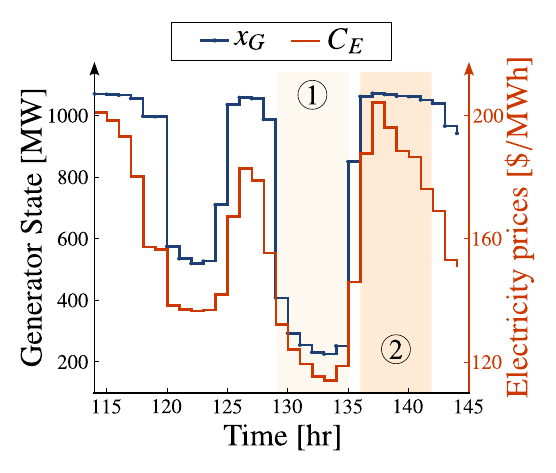}
    \caption{Optimal generator state.}
    \label{subfig:G_state_C1}
    \end{subfigure}%
    \begin{subfigure}{0.25\textwidth}
    \centering
    \includegraphics[scale=0.49]{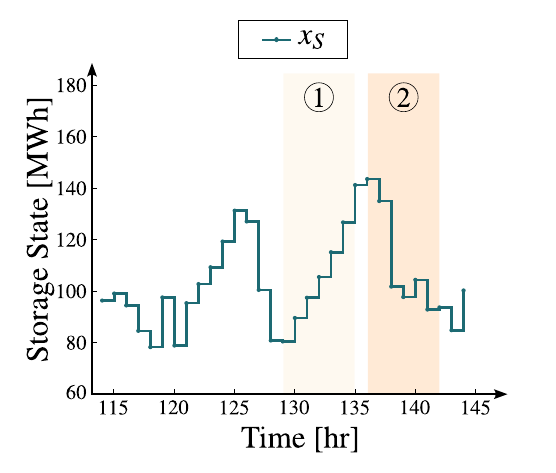}
    \caption{Optimal storage state.}
    \label{subfig:S_state_C1}
    \end{subfigure}%
    \begin{subfigure}{0.25\textwidth}
    \centering
    \includegraphics[scale=0.49]{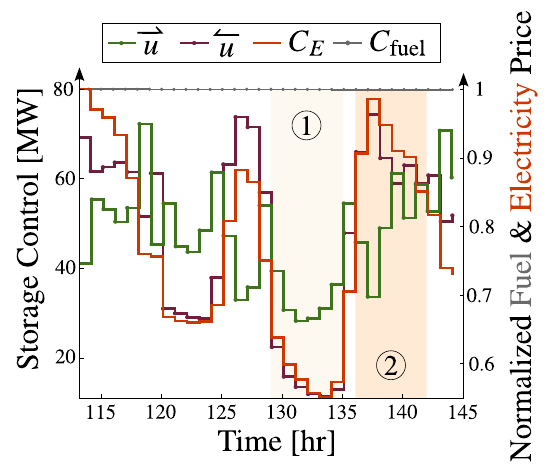}
    \caption{Optimal storage control.}
    \label{subfig:S_control_C1}
    \end{subfigure}%
    \begin{subfigure}{0.25\textwidth}
    \centering
    \includegraphics[scale=0.49]{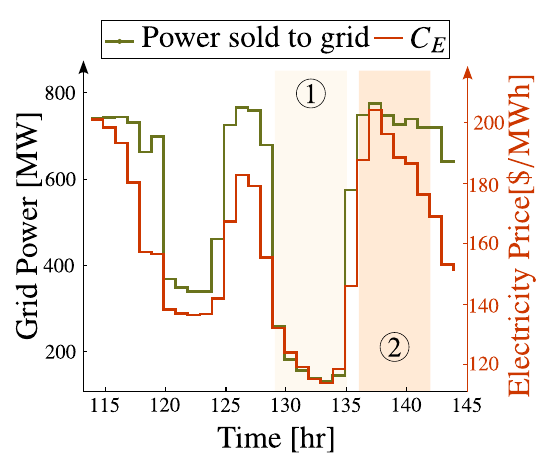}
    \caption{Optimal power to grid.}
    \label{subfig:Grid_C1}
    \end{subfigure}%
    \captionsetup[figure]{justification=centering}
    \caption{Case Study I: Optimal state and control variables for combined cycle generator with thermal storage and carbon capture. Here $x_{G}$ and $x_{S}$ are generator and storage states, respectively; $\protect\overrightharpoonsub{u}{}$, and $\protect\overleftharpoonsub{u}{}$ are charge and discharge control signals; $C_E$ and $C_{\text{fuel}}$ are electricity and fuel price signals.} 
    \label{fig:Results_C1}
\end{figure*}

\begin{figure}[t]
    \centering
    \includegraphics[scale=0.74]{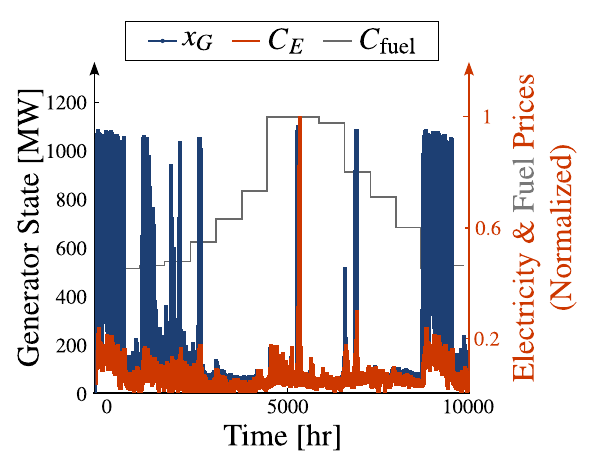}
    \caption{Case Study I: Optimal generator state ($x_{G}$) with electricity and fuel prices ($C_E$, and $C_{\text{fuel}}$) over a long horizon of around 42 days.}
    \label{subfig:G_state_C1_long}
\end{figure}

The problem is carried out for $30$ years of operation with an hourly time mesh (262980 time grid points), and it is assumed that thermal energy can not be directly sold; rather, it can only be used for meeting the thermal load demand from CCS.
The input parameters associated with this case study, which are largely based on Refs.~\cite{Vercellino2022}--\cite{energy2020capital} and are tabulated in Tables~\ref{tab:par1} and \ref{tab:CaseIPar}.
Due to linearity and convexity, a globally optimal solution is found efficiently in only $280~[\unit{s}]$.
The results from this case study are discussed in the following and shown in Figs.~\ref{fig:Results_C1} and \ref{subfig:G_state_C1_long}.

The behavior of the candidate generator and storage system within the optimization problem is tightly associated with the cost of fuel and electricity.
This study assumes that fuel prices are constant within each month.
To understand the impact of the electricity price signal on the system's behaviors, we first interpret the results within a single month with fixed fuel prices.
We have marked Fig.~\ref{fig:Results_C1}, with time periods \circled{1} and \circled{2}, which are associated with low and high electricity prices, respectively.

Accordingly, as shown in Fig.~\ref{subfig:G_state_C1}, 
in the region marked by \circled{1}, electricity prices are low.
Thus, operating the generator at full capacity is not profitable.
However, this time window is a good opportunity for the generator to charge the storage, as shown in Fig.~\ref{subfig:S_state_C1}.
Figure~\ref{subfig:S_control_C1} shows the charge and discharge signals during this period.
Note that while both charging and discharging signals are active, the charging signal is larger, resulting in an increase in the storage state.
During this period, the power sold to the grid is relatively lower than in other time periods, as shown in Fig.~\ref{subfig:Grid_C1}.

The region marked by \circled{2} describes the system's response to a scenario in which the electricity prices are relatively high.
During this period, it is profitable for the generator to run at full or high capacity (see Fig.~\ref{subfig:G_state_C1}).
According to Figs.~\ref{subfig:S_state_C1} and \ref{subfig:S_control_C1}, high electricity prices also incentivize storage discharge during this period.
The discharged thermal energy during this phase removes the dependent thermal load from the generator, allowing an increase in the amount of electricity sold to the grid.
This is shown in Fig.~\ref{subfig:Grid_C1}.

To understand how the system behavior is affected by the combination of electricity and fuel prices, a longer horizon, during which both price signals change, is presented in Fig.~\ref{subfig:G_state_C1_long}.
from this figure, it is clear that when the electricity prices are low and the fuel prices are high, the operation of the generator drops significantly to about $10\%$ of its nominal capacity. 
This figure provides insights into the impact of fuel and electricity prices on the system's behavior, incentivizing the inclusion of future price increases (due to inflation, etc.) in our future work.  

Overall, the optimal storage capacity is found to be $237.53~[\unit{MWh}]$, and the NPV objective is $\$-1.83\times 10^{9}$. 
Compared to the case with no thermal storage available (with CCS), this solution represents a $71.03\%$ improvement in the project's economic viability.
These results use several critical assumptions that should be reconsidered for evaluating a specific site location.
Furthermore, current historical market signals were extrapolated from \cite{gridstatus, NG}, but including predicted future market signals (e.g., from Ref.~\cite{JesseDJenkins2021, GenX}) could aid in the decision marking by considering the future grid composition and climate change.
Even with these current assumptions, the expected trends are observed, and \ToolName{} can be used as a strong framework for the early-stage investigation of these generator and storage technologies. 

\begin{figure}[t]
    \captionsetup[subfigure]{justification=centering}
    \centering
    \begin{subfigure}{0.5\columnwidth}
    \centering
    \includegraphics[scale=0.7]{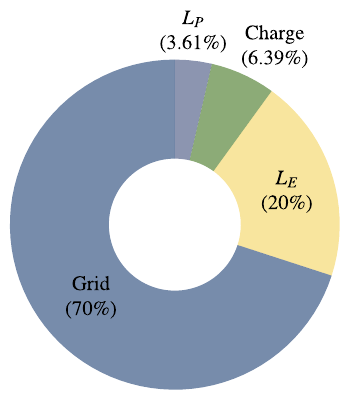}
    \caption{Generator energy.}
    \label{subfig:Pie_Gen_C_1}
    \end{subfigure}%
    \begin{subfigure}{0.5\columnwidth}
    \centering
    \includegraphics[scale=0.7]{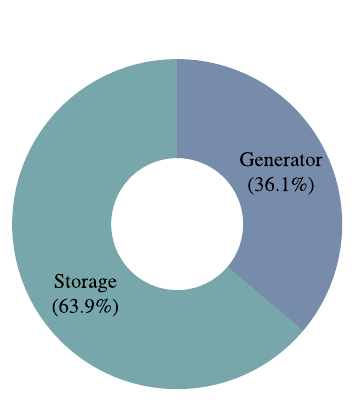}
    \caption{Primary load.}
    \label{subfig:Pie_Primary_C_1}
    \end{subfigure}%
    
    \begin{subfigure}{0.5\columnwidth}
    \centering
    \includegraphics[scale=0.7]{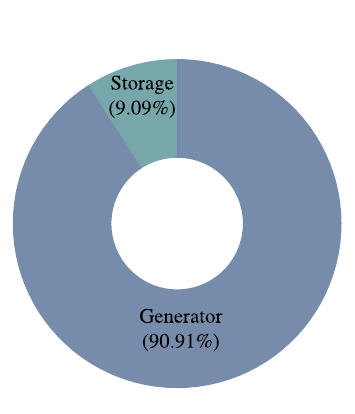}
    \caption{Revenue.}
    \label{subfig:Pie_Rev_C_1}
    \end{subfigure}%
    \captionsetup[figure]{justification=centering}
    \caption{Breakdown of various elements in Case Study \rom{1}, a combined cycle with thermal storage and carbon capture: (a) Generator energy usage, (b) Primary load contributions, and (c) Revenue contributions.}
    \label{fig:Pie_C_1}
\end{figure}

Assessing the share of various functions and requirements over the plant's lifetime can yield insights regarding its optimal operation.
Specifically, Fig.~\ref{fig:Pie_C_1} shows how the energy produced by the generator is utilized.
Specifically, $70\%$ of the generator energy is sold to the electricity grid, while $20\%$, $6.39\%$, and $3.61\%$ are used for electric load, charge, and primary load, respectively.
The relatively small share of the generator in satisfying the primary load is made possible through the inclusion of the thermal energy system, which is responsible for satisfying $63.9\%$ of the thermal energy demand from CCS according to Fig.~\ref{subfig:Pie_Primary_C_1}.
As shown in Fig.~\ref{subfig:Pie_Rev_C_1}, the storage system is responsible for about $9.09\%$ of the overall revenue.

NGCC generators have traditionally served as peaking power plants, meaning that they are used during the peak electricity demand to satisfy the load.
The requirement for following some electricity demand or the commitment to provide a certain amount of power to the grid during specific time periods has direct implications on optimal system behavior and storage sizing.
To assess this impact, we consider the case in which the power plant is required to operate at its full capacity during peak hours $T_{\text{peak}}$.
Moreover, since the generated power during this time period is committed to the electricity grid, no storage charging is allowed. 
These considerations can be included in the optimization problem by updating some of the bounds in Eqs.~(\ref{eq:discharge})--(\ref{eq:controlcons2}).
The new operational constraints are mathematically formulated as:
\begin{align}
\label{eq:NGCC_Demand_Following_gen}
\overrightharpoonsub{u}{\text{max}}(t) & = 
            \begin{cases}
    		\bm{0}, & \text{if~} t \in T_{\text{peak}} \\
            \overrightharpoonsub{u}{\text{max}}(t), & \text{otherwise}
            \end{cases}\\
\bm{x}_{G,\text{min}}(t) & =
            \begin{cases}
			\bm{x}_{G,\text{max}}(t), & \text{if~} t \in T_{\text{peak}} \\
            \bm{0}, & \text{otherwise}
		      \end{cases}
\end{align}

Assuming the daily peak power demand of $T_{\text{peak}} = [3-7]~\text{PM}$, we solve the Case Study \rom{1} in the presence of the demand-following constraint. 
As opposed to the previous case, where the generator would shut down or operate at a lower capacity when electricity prices were low and fuel costs were high, this case demonstrates a situation in which the NGCC power plant operates in a more restricted market. 
Compared to the previous case, the NGCC commitment to provide power to the grid regardless of electricity and fuel prices during the peak demand period negatively affects the economics of the plant, changing the NPV from $\$-1.83 \times 10^9$ to $\$-3.91 \times 10^9$.

This scenario motivates more storage activity compared to the previous case.
Specifically, since no charging is allowed during the peak demand, the overall charging activity is increased during non-peak hours when compared to the previous case. 
More extreme discharge activity is witnessed during the peak electricity demand, particularly during times when electricity prices are low and fuel costs are high.
The discharge signal contributes to meeting the thermal load requirements for CCS operation, which is largest when the power plant is operating at its full capacity during peak demand hours. 
The important role of storage in this system is demonstrated by the $40.56\%$ increase in optimal storage capacity from $237.53~[\unit{MWh}]$ to $333.87~[\unit{MWh}]$.

\begin{figure}[t]
    \centering
    \includegraphics[scale=0.74]{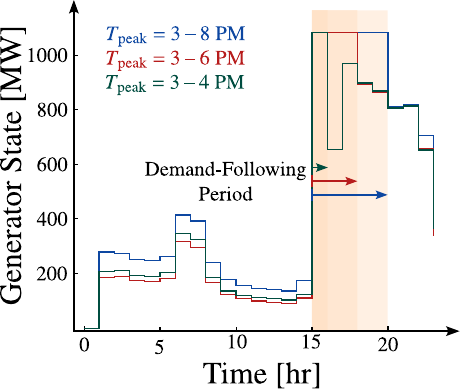}
    \caption{Generator state with demand-following constraints with $3$ different demand-following periods.}
    \label{fig:sentraj}
\end{figure}

\begin{figure}[t]
    \captionsetup[subfigure]{justification=centering}
    \centering
    \begin{subfigure}{0.5\columnwidth}
    \centering
    \includegraphics[width = \columnwidth]{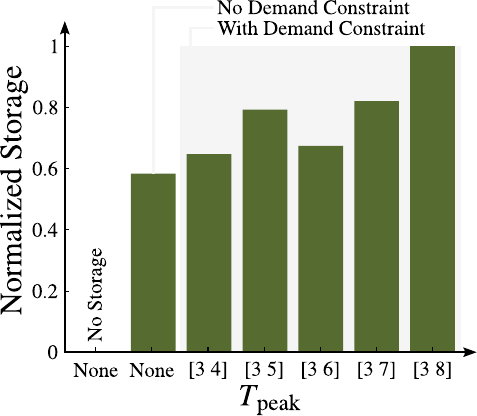}
    \caption{Optimal storage capacity.}
    \label{subfig:CaseI_sens_P}
    \end{subfigure}%
    \begin{subfigure}{0.5\columnwidth}
    \centering
    \includegraphics[width = \columnwidth]{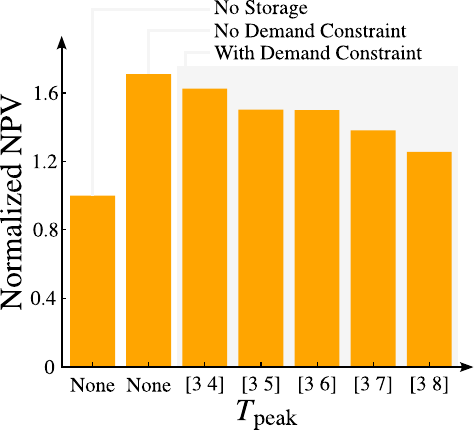}
    \caption{Optimal NPV.}
    \label{subfig:CaseI_sens_F}
    \end{subfigure}%
    \captionsetup[figure]{justification=centering}
    \caption{Case Study \rom{1}: Sensitivity of normalized optimal storage capacity and net present value (NPV) objective function to the duration of the demand-following constraint for the natural gas combined cycle (NGCC) power plant with carbon capture and storage (CCS) and thermal energy storage (TES).}
    \label{fig:CaseI_sense}
\end{figure}

\begin{table}[t]
    \centering
    \caption{Sensitivity of the optimal storage capacity and NPV objective function to variations in the duration of the demand-following constraint for Case Study \rom{1}.}
        \label{tab:CaseIsens}
    \renewcommand{\arraystretch}{1.1}
    \begin{tabular}{c| c c}
        \hline  \hline
         \textbf{Scenario} & \textbf{Optimal Storage} & \textbf{Optimal NPV} \\
         \hline
          $\text{No Storage with CCS}$ & -  & $-6.3170$ \\
          $\text{No demand constraint }$ & $237.5290$ & $-1.8298$ \\
          $T_{\text{peak}} \in [3-4]~\text{PM}$ & $263.5106$ & $-2.3677$ \\
          $T_{\text{peak}} \in [3-5]~\text{PM}$ & $322.4310$ & $-3.1409$ \\
          $T_{\text{peak}} \in [3-6]~\text{PM}$   & $274.4213$ & $-3.1627$ \\
          $T_{\text{peak}} \in [3-7]~\text{PM}$  & $333.8728$ & $-3.9114$ \\
          $T_{\text{peak}} \in [3-8]~\text{PM}$  & $406.8284$ & $-4.7072$ \\
         \hline
    \end{tabular}
\end{table}

The sensitivity of the NPV objective function and storage capacity to the duration of the demand-following constraint is discussed next.
Here, we perturb the duration of the demand-following constraint and examine the variations in optimal system behavior. 
Figure~\ref{fig:sentraj} presents the generator state in the presence of the demand-following constraint for $3$ different demand-following periods.  
To facilitate comparisons, both NPV and storage capacity are normalized according to the optimal results with no demand-following constraint.
The results from this study are tabulated in Tab.~\ref{tab:CaseIsens} and shown in Fig.~\ref{fig:CaseI_sense}.
According to these results, it is clear that under the current assumptions (HES architecture, carbon tax, price signals, parameters, etc.), the optimal thermal energy storage capacity generally increases with increasing the demand-following duration, with the exception of the results from $T_{\text{peak}} = [3~6]$.
The optimal NPV decreases with increasing the demand duration.

Two important conclusions can be made by analyzing the optimal system behavior: (i) In a more restricted market scenario characterized by the presence of demand-following constraints, the role of the thermal energy storage system becomes increasingly prominent. (ii) Operating in an increasingly restricted market often results in poor economic performance of the integrated energy system.
As is evident from these studies, \ToolName{} enables the optimal performance assessment of the integrated energy system, facilitating the decision-making process in early-stage design with consideration of different scenarios, policies, and requirements.

\subsection{Case Study \rom{2}: Wind Farm with Battery Energy Storage Units} \label{subsec:CaseStudy2}
In this case study, we consider an on-shore wind farm with a large plant footprint operating at $200~[\unit{MW}]$ in the Great Plains region in combination with a $50~[\unit{MW}]$ battery storage rate.
These ratings were selected because of the availability of reliable technical and economic data \cite{energy2020capital}.%
The wind farm parameters are based on a case with $71$ wind turbines, each with a nominal capacity of $p_{\text{rated}} = 2.8~[\unit{MW}]$, with a rotor diameter of $ D = 125~[\unit{m}]$ and a hub height of $90~[\unit{m}]$ \cite{energy2020capital}.

Variations in wind speed cause fluctuations in the power outputted to the grid, negatively affecting power quality and stability.
Therefore, a BESS is needed to mitigate these adverse impacts \cite{JANNATI}. 
Lithium-ion batteries offer high power density and energy efficiency, making them a promising energy storage technology for wind farm applications \cite{Jiang}.
Therefore, a utility-scale lithium-ion battery consisting of $25$ modular, prefabricated battery storage container buildings is considered here \cite{energy2020capital}.
Further details about the parameters used here can be found in Ref.~\cite{energy2020capital}.
To better highlight the value of the proposed framework for future planning exercises, $C_\text{occ}$ associated with BESS reported in Tab.~\ref{tab:par1} is initially selected to be $50\%$ of the cost reported in Ref.~\cite{energy2020capital}.
\begin{figure*}[t]
    \captionsetup[subfigure]{justification=centering}
    \centering
    \begin{subfigure}{0.25\textwidth}
    \centering
    \includegraphics[scale=0.49]{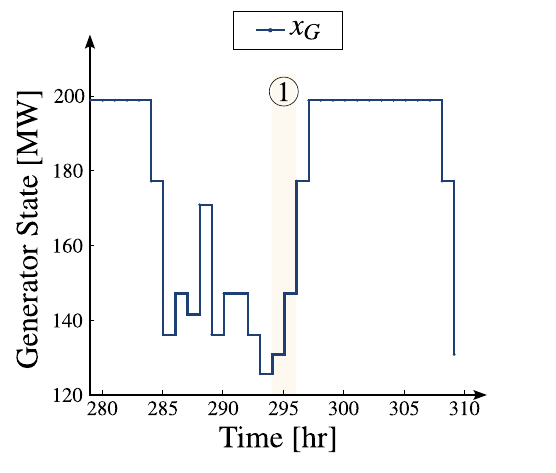}
    \caption{Optimal generator state.}
    \label{subfig:G_state_C2}
    \end{subfigure}%
    \begin{subfigure}{0.25\textwidth}
    \centering
    \includegraphics[scale=0.49]{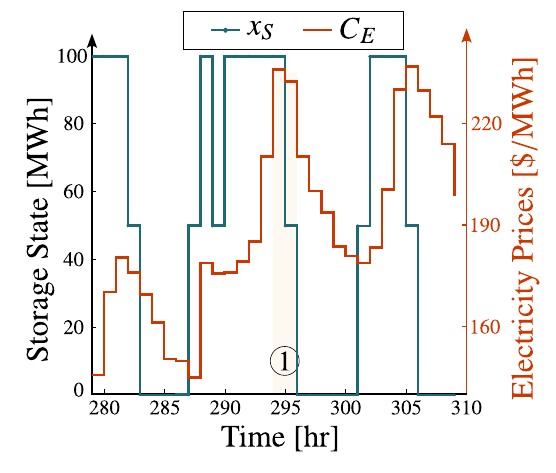}
    \caption{Optimal storage state.}
    \label{subfig:S_state_C2}
    \end{subfigure}%
    \begin{subfigure}{0.25\textwidth}
    \centering
    \includegraphics[scale=0.49]{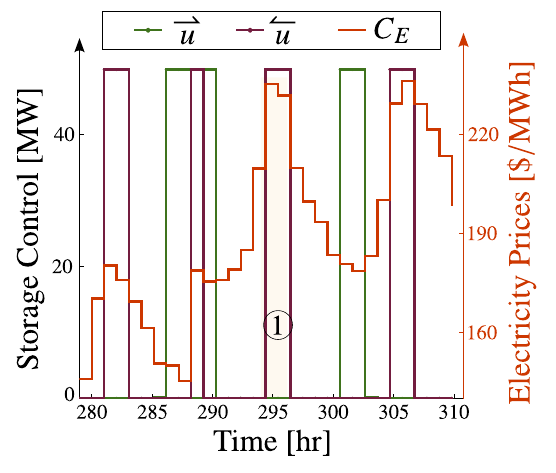}
    \caption{Optimal storage control.}
    \label{subfig:S_control_C2}
    \end{subfigure}%
    \begin{subfigure}{0.25\textwidth}
    \centering
    \includegraphics[scale=0.49]{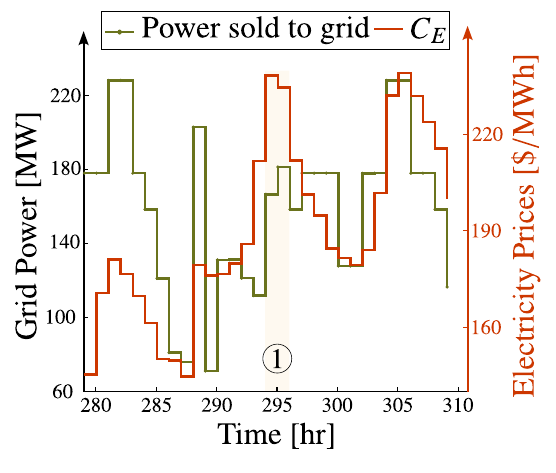}
    \caption{Optimal power to grid.}
    \label{subfig:Grid_C2}
    \end{subfigure}%
    \captionsetup[figure]{justification=centering}
    \caption{Case Study II: Optimal state and control variables for a wind farm with a battery storage unit. Here $x_{G}$ and $x_{S}$ are generator and storage states, respectively; $\protect\overrightharpoonsub{u}{}$, and $\protect\overleftharpoonsub{u}{}$ are charge and discharge control signals; $C_E$ and $C_{\text{fuel}}$ are electricity and fuel price signals.}
    \label{fig:Results_C2}
\end{figure*}

As opposed to Case Study I, where the operator can request a specific power output from the generator, the wind farm operation is primarily determined by the availability of wind.
Using wind speed $v_{w}(t)$ as an additional time-dependent input to the model, the upper bound of Eq.~(\ref{eq:controlcons2}) was estimated for each turbine based on the wind speeds, wind farm specifications, and the capacity factor of $c_{p}=0.55$:
\begin{align}
    p_{w} = c_{p}\frac{1}{2}\rho_{air}\frac{\pi D^2}{4}v_{w}(t)^3
\end{align}

\noindent
where $p_{w}$ is the wind power, and $\rho_{air}$ is the air density. 
The maximum wind power is assumed to happen at $v_{w} = 25~[\unit{m/s}]$, and the turbine is off for wind speeds above this value. 
All of the power vector elements greater than $p_{\text{rated}}$ are saturated at this value.
Finally, the available power output takes into consideration the number of wind turbines in the farm. 
The input parameters associated with this case study are tabulated in Tables~\ref{tab:par1} and \ref{tab:CaseIIPar}.
The candidate system is illustrated in Fig.~\ref{fig:IES_CaseII}, where the electrical load is present due to the need to operate auxiliary equipment in the facility (auxiliary loads).

\begin{table}[t]
    \centering
    \caption{Parameters associated with Case Study \rom{2} for a wind farm connected to a battery energy storage system, largely based on Ref.~\cite{energy2020capital}.}
        \label{tab:CaseIIPar}
    \renewcommand{\arraystretch}{1.1}
    \begin{tabular}{r| r r r| r r}
        \hline  \hline
         \textbf{Field} & {\textbf{Value}} & \textbf{Unit} & \textbf{Field} & {\textbf{Value}} & \textbf{Unit}\\
         \hline
          $\rho_{\text{fuel}}$ & $0$ & \unit{kg/h.MW} & $\alpha_{\text{CO\textsubscript{2}}}$ & $0$ & \unit{ton/kg}  \\
          $\tau$ & $0$ & \unit{h} & $\eta_{G}$ & $1$ & - \\
          $u_{G,\text{min}}$ & $0$ & \unit{MW} & $u_{G,\text{max}}$ & $200$ & \unit{MW}\\
          $x_{G,\text{min}}$ & $0$ & \unit{MW}& $x_{G,\text{max}}$ & $200$ & \unit{MW}\\
          $x_{S}(t_{0})$ & $0$ & \unit{MWh}& $x_{S}(t_{f})$ & $0$ & \unit{MWh}\\
          $\overrightharpoonsub{u}{\text{max}}$ & $50$ & \unit{MW} & $\overleftharpoonsub{u}{\text{max}}$ & $50$ & \unit{MW}\\
          $L_{P}$ & $0$ & \unit{MW} & $L_{E}$ & $0.1 x_{G}$ & \unit{MW}\\
         $T_{\text{con}}$ & $5$ & \unit{years} & $r$ & $0.075$ & -\\
         \hline
    \end{tabular}
\end{table}

Unlike in the previous case, the generator power level here is mainly determined by the availability of wind and wind turbine characteristics.
Therefore, in Fig.~\ref{subfig:G_state_C2}, the generator captures all the wind power that it is capable of harvesting.
On the other hand, the charge and discharge decisions are heavily determined by the electricity prices.
As it is clear from Fig.~\ref{subfig:S_control_C2}, when the electricity prices are high, the charge signal turns off, and the discharge signal activates.
This results in dramatic drops in the storage state, as evidenced by region \circled{1}, in Figs.~\ref{subfig:S_state_C2} and \ref{subfig:S_control_C2}.
Ideally, this outcome should increase the power sold to the grid and, thus, increase revenue.
However, since the wind power available in this period is relatively lower than the neighboring time periods (see Fig.~\ref{subfig:G_state_C2}), the power sold to the grid remains comparatively at a medium level.
This is shown in Fig.~\ref{subfig:Grid_C2}.

\begin{figure}[t]
    \captionsetup[subfigure]{justification=centering}
    \centering
    \begin{subfigure}{0.5\columnwidth}
    \centering
    \includegraphics[scale=0.7]{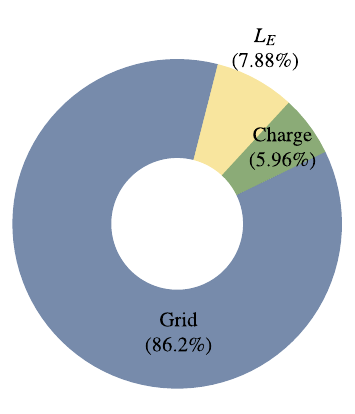}
    \caption{Generator energy.}
    \label{subfig:Pie_Gen_C_2}
    \end{subfigure}%
    \begin{subfigure}{0.5\columnwidth}
    \centering
    \includegraphics[scale=0.7]{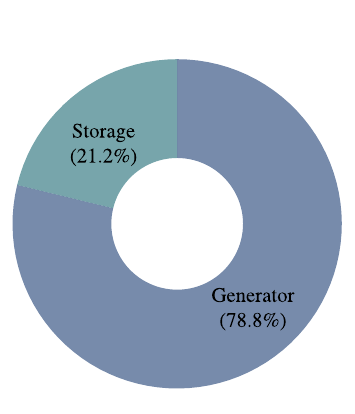}
    \caption{Electrical load.}
    \label{subfig:Pie_Primary_C_2}
    \end{subfigure}%
    
    \begin{subfigure}{0.5\columnwidth}
    \centering
    \includegraphics[scale=0.7]{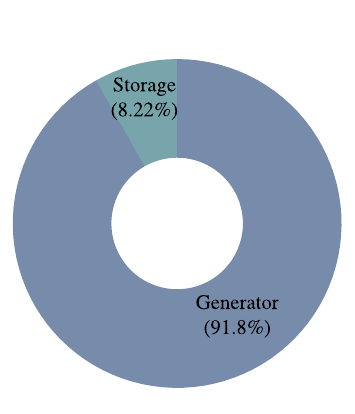}
    \caption{Revenue.}
    \label{subfig:Pie_Rev_C_2}
    \end{subfigure}%
    \captionsetup[figure]{justification=centering}
    \caption{Breakdown of various elements in Case Study \rom{2}, a wind farm with battery energy storage units: (a) Generator energy usage, (b) Electrical load contributions, and (c) Revenue contributions.}
    \label{fig:Pie_C_2}
\end{figure}

The optimal storage capacity is $100~[\unit{MWh}]$, equivalent to $2$ hours of storage, and the NPV objective function is $\$~0.74\times10^{9}$, indicating that for a large wind farm, a utility-scale BESS has the potential to be economically competitive under these assumptions.
The percentages associated with the energy produced by the generator and the shares of the generator and storage unit from the electrical load and revenue are shown in Fig.~\ref{fig:Pie_C_2}. 
Accordingly, $5.96\%$ of the generated energy is used to charge the BESS, while $5.96\%$ is used to satisfy the electric load demands. 
Over the project's life, the storage will satisfy $21.2\%$ of this load demand, and the generator will satisfy the remainder.
Finally, storage contributions to the revenue are estimated as $8.22\%$.

\begin{figure*}[t]
    \captionsetup[subfigure]{justification=centering}
    \centering
    \begin{subfigure}{0.5\textwidth}
    \includegraphics[scale=0.74]{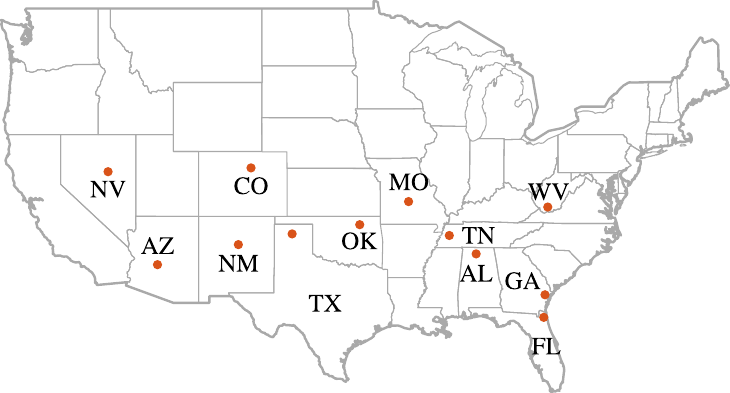}
    \caption{Candidate farm locations.}
    \label{fig:US_Locations}
    \end{subfigure}%
    \begin{subfigure}{0.5\textwidth}
    \centering
    \includegraphics[scale= 0.74]{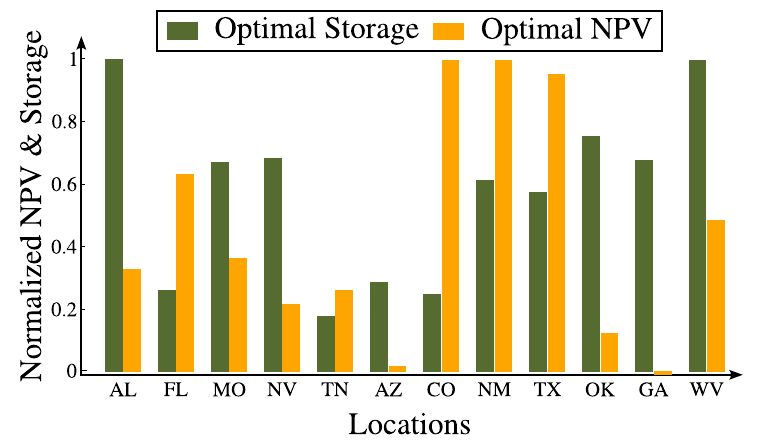}
    \caption{Optimal storage capacity and NPV.}
    \label{subfig:Locations_storage_NPV_wind}
    \end{subfigure}%
    \captionsetup[figure]{justification=centering}
    \caption{Case Study \rom{2}: Optimal storage capacity and NPV objective for the wind farm with a battery energy storage system (BESS) at various U.S. locations.}
    \label{fig:Locations_wind}
\end{figure*}

Note that the optimal storage sizing in this study is strongly dependent on costs associated with the storage facility, among which the overnight capital cost is a dominant parameter.
To examine the impact of this parameter, an additional study were solved in which $C_{\text{occ}}$ was increased to $75\%$ the value from Ref.~\cite{energy2020capital}, resulting in a decrease in the optimal battery storage sizing to $50~[\unit{MWh}]$ (i.e., an hour of storage with the current charge/discharge rate) from $100~[\unit{MWh}]$ in the original study.
If we use $C_{\text{occ}}$ from Ref.~\cite{energy2020capital} directly (i.e., $100\%$), we note that a BESS is no longer an economically viable decision for this HES.  
These investigations highlight the value of the proposed framework in not only optimally sizing the storage but also determining if integrating a storage system is economically viable.

Among others, the solution presented here strongly depends on the selected wind profile.
The location of the farm, however, has a significant impact on its technical and economic performance \cite{Azads2024Acc}.
To gain a deeper understanding of the impact of wind profile and its tight association with the location of the farm, we extend this study to assess the performance of the farm across various U.S. locations, using the average hourly wind data collected and maintained by the National Oceanic and Atmospheric Administration in Ref.~\cite{WindData}.
While interval prediction can be used to predict wind power output scenarios \cite{Luo}, nominal average hourly values based on historical data are used here. 
The candidate locations for the wind farm are shown in Fig.~\ref{fig:US_Locations}.
The NPV objective function, along with the optimal sizing of the BESS at each location, is found and presented in Fig.~\ref{subfig:Locations_storage_NPV_wind}.

These investigations are performed in the presence of a multitude of assumptions, including fixed electricity prices across all states.
From these figures, it is clear that wind farms located in Colorado (CO), New Mexico (NM), and Texas (TX) have a high potential for successful commercialization. 
These three site locations are characterized by high winds (and thus, high mean power output from the farm) with a small standard deviation in the generated wind power.
The relatively high and constant availability of wind energy in these regions informs the need for relatively small BESS under the considered electricity price signal. 
The largest optimal storage capacity belongs to Alabama (AL) and West Virginia (WV).
Both of these site locations are characterized by medium mean wind farm power and relatively high standard deviation.
Note that while wind profile characteristics are determining factors in assessing the performance of the wind farm, their variation in relation to the electricity prices is the driver behind the optimal results presented here.  
Such complex interactions necessitate the need for an efficient optimization-based framework, such as \ToolName{}, to inform optimal decision-making in defining crucial technical aspects of an HES.

\subsection{Case Study \rom{3}: Nuclear Power Plant with Hydrogen Generation}
\label{subsec:CaseStudy3}

Nuclear power plants (NPPs) play a crucial role in the future of the electricity market, as after hydropower, they are the largest energy source with low carbon emissions \cite{BPStatistical}.
Nevertheless, their economic viability is challenged by low electricity prices from other generation sources, as well as other complexities.
For example, the most expensive per unit electricity, which is typically produced by nuclear power (due to complexity, capital intensiveness, construction time, etc.), is purchased last during peak demand when other sources are not available.
This, along with other factors such as costs of repair, has resulted in the retirement of over 12 nuclear power plants (NPP) from February 2013 through April 2021 \cite{HERON}.
The retirement of NPPs, which roughly produce a fifth of the total electricity generation and half of the non-fossil fuel-based electricity in the U.S. (with no intermittency), points to a changing energy landscape in which NPPs require flexibility in base load to remain competitive \cite{McDowell2021}.

A potential solution is to increase the economic competitiveness of NPPs by configuring them as a part of hybrid energy systems, operating simultaneously with other energy technologies.  
This strategy will enable NPPs to supply power to the grid when electricity prices are high and focus on storage during periods of oversupply.
Depending on the storage type, the stored energy can then be directly sold as electricity or in the form of another energy/commodity, such as thermal energy or hydrogen.
The thermal energy can be directly sold to chemical plants for use in industrial processes, while hydrogen can be either combusted to generate additional electricity \cite{Khamis2010} or sold for use in fuel cells or the steel manufacturing industry.
The growing market for hydrogen \cite{IEA_Hydro} signifies a unique opportunity for NPPs to expand into additional markets since the integration of NPPs with additional markets and technologies has the potential to keep NPPs competitive \cite{Hancock2022, Zhang2022, Frick2019}.

Therefore, this case study considers the integration of NPP with a hydrogen facility and storage.
This hybrid operation \cite{Frick2019} releases NPPs from their traditional baseload by enabling them to strategically (based on economics) produce hydrogen or sell electricity to the grid.
Here, we consider an NPP with two pressurized water reactors (PWR), in which the heat generated by the fuel in the reactor is released into the surrounding pressurized cooling water.
The pressurized water absorbs the heat without boiling and, after passing through a steam generator, flows through a steam turbine to generate electricity.  
Hydrogen is produced via the high-temperature steam electrolysis (HTSE) process \cite{Mingyi2008}, which requires both thermal and electrical connections to the NPP.
Based on load requirements reported in Ref.~\cite{Frick2019}, this study assumes that for every unit of electricity, $10\%$ thermal energy is required to produce hydrogen.
The thermal requirement is then defined as a function of the input electricity from the tertiary charging signal, effectively ensuring that the thermal loads are only present when a decision to produce hydrogen is made.  
The cost of the fuel is constructed based on predictions presented in Ref.~\cite{Kryzia2016}, while hydrogen market prices are assumed to be fixed at $\$7.0$ per kg of hydrogen, which is within the price range reported in Ref.~\cite{Ramadan2022}.  
The cost parameters for the NPP and the hydrogen generation and storage facility, which are largely based on Refs.~\cite{energy2020capital, Frick2019}, are described in Tab.~\ref{tab:par1}. 
The remaining problem parameters are shown in Tab.~\ref{tab:CaseIIIPar}.

Informed by hydrogen market operations and the fact that sales of hydrogen can only occur at pre-scheduled times, this case study considers a discrete daily demand through the following constraint:
\begin{align}
\label{eq:NPP_rev_cons}
{0} \leq {u}_{TR}(t) \leq
\begin{cases}
{u}_{\text{max}}, & \text{if~} t = 8 ~\text{AM} \\
{0}, & \text{otherwise}
\end{cases}
\end{align}

\noindent
According to this equation, hydrogen sales can only occur between $8-9~\text{AM}$ on a daily basis.

\begin{table}[ht]
    \centering
    \caption{Parameters associated with Case Study \rom{3} for a nuclear power plant connected to a hydrogen production and storage facility, largely based on Refs.~\cite{energy2020capital,Frick2019, Settle2009, OECDrep, Hancock2022}.}
        \label{tab:CaseIIIPar}
    \renewcommand{\arraystretch}{1.1}
    \begin{tabular}{r| r r r| r r}
        \hline  \hline
         \textbf{Field} & {\textbf{Value}} & \textbf{Unit} & \textbf{Field} & {\textbf{Value}} & \textbf{Unit}\\
         \hline
          $\rho_{\text{fuel}}$ & $0.001$ & \unit{kg/h.MW} & $\alpha_{\text{CO\textsubscript{2}}}$ & $0$ & \unit{ton/kg}  \\
          $\tau$ & $1.79$ & \unit{s} & $\eta_{G}$ & $1$ & - \\
          $u_{G,\text{min}}$ & $0$ & \unit{MW} & $u_{G,\text{max}}$ & $2156$ & \unit{MW}\\
          $x_{G,\text{min}}$ & $0$ & \unit{MW}& $x_{G,\text{max}}$ & $2156$ & \unit{MW}\\
          $x_{S}(t_{0})$ & $50$ & \unit{kg}& $x_{S}(t_{f})$ & $50$ & \unit{kg}\\
          $\overrightharpoonsub{u}{\text{max}}$ & $1065$ & \unit{MW} & $\overleftharpoonsub{u}{\text{max}}$ & $27990$ & \unit{kg/h}\\
          $L_{P}$ & $0.1 \overrightharpoon{u} $ & \unit{MW} & $L_{E}$ & $0.1 x_{G}$ & \unit{MW}\\
         $T_{\text{con}}$ & $7$ & \unit{years} & $r$ & $0.075$ & -\\
         $\alpha_{ET}$ & $0.0377$ & \unit{MWh/kg} & $\alpha_{TE}$ & $29.762$ & \unit{kg/MWh}  \\ 
         \hline
    \end{tabular}
\end{table}

\begin{figure*}[t]
    \captionsetup[subfigure]{justification=centering}
    \centering
    \begin{subfigure}{0.25\textwidth}
    \centering
    \includegraphics[scale=0.49]{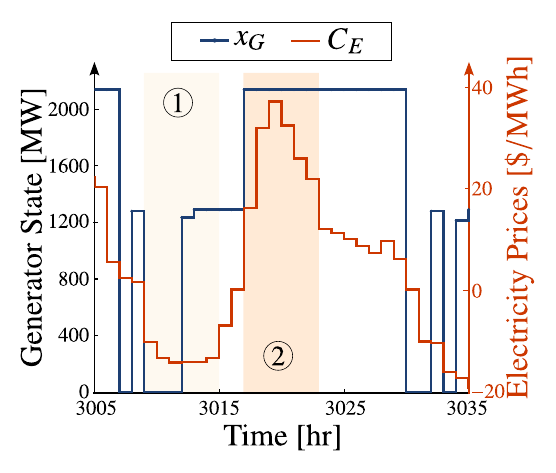}
    \caption{Optimal generator state.}
    \label{subfig:G_state_C3}
    \end{subfigure}%
    \begin{subfigure}{0.25\textwidth}
    \centering
    \includegraphics[scale=0.49]{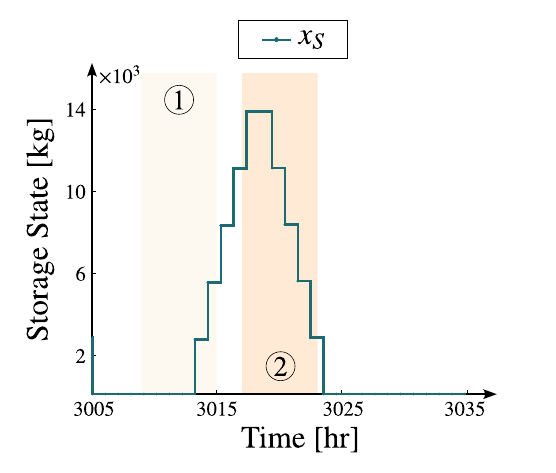}
    \caption{Optimal storage state.}
    \label{subfig:S_state_C3}
    \end{subfigure}%
    \begin{subfigure}{0.25\textwidth}
    \centering
    \includegraphics[scale=0.49]{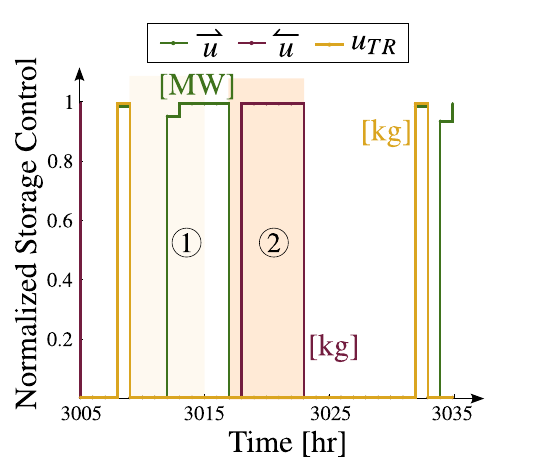}
    \caption{Optimal storage control.}
    \label{subfig:S_control_C3}
    \end{subfigure}%
    \begin{subfigure}{0.25\textwidth}
    \centering
    \includegraphics[scale=0.49]{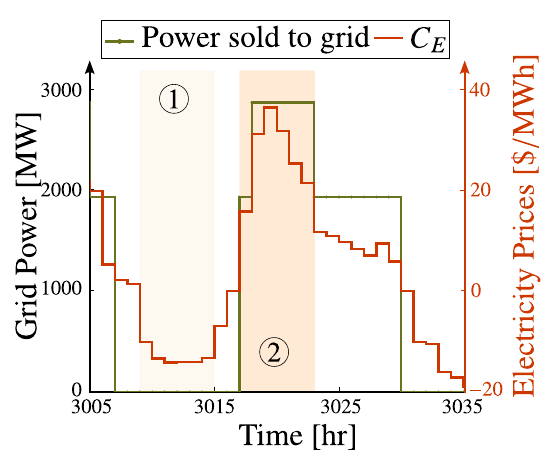}
    \caption{Optimal power to grid.}
    \label{subfig:Grid_C3}
    \end{subfigure}%
    \captionsetup[figure]{justification=centering}
    \caption{Case Study III: Optimal state and control variables for a nuclear power plant with a hydrogen production and storage facility.
    Here $x_{G}$ and $x_{S}$ are generator and storage states, respectively; $\protect\overrightharpoonsub{u}{}$, and $\protect\overleftharpoonsub{u}{}$ are charge and discharge control signals; $C_E$ and $C_{\text{fuel}}$ are electricity and fuel price signals.}
    \label{fig:Results_C3}
\end{figure*}

The problem is solved for $30$ years of operation, and the results are presented in Fig.~\ref{fig:Results_C3}. 
According to Fig.~\ref{subfig:G_state_C3}, the NPP power level is initially zero and then starts to increase in in region \circled{1}, where the electricity prices are low.
In region \circled{2}, where electricity prices are high the NPP power level increases to full capacity.
As evidenced by Fig.~\ref{subfig:S_state_C3}, during phase \circled{1} and \circled{2}, the level of hydrogen storage increases and decreases, respectively. 
These incremental changes are directly influenced by the charge and discharge rates, which are currently among the problem parameters, but could also be investigated as plant optimization variables, similar to Ref.~\cite{Vercellino2022}.
According to Fig.~\ref{subfig:S_control_C3}, direct sales of hydrogen, which become possible every day between $8-9~\text{AM}$, result in a periodic discharge of hydrogen, in which the discharged hydrogen is directly sold to the market.
Note that in this figure, the discharge signal is mostly hidden behind the revenue signal.
The only exception is the full discharge between $t\in[3018, 3022]$ in which hydrogen is combusted to generate electricity and revenue.
The power sold to the grid is shown in Fig.~\ref{subfig:Grid_C3}, where it is evident that the power sold to the grid is the highest in region \circled{2} when the electricity prices are high.

\begin{figure}[t]
\centering
    \captionsetup[subfigure]{justification=centering}
    \centering
    \begin{subfigure}{0.5\columnwidth}
    \centering
    \includegraphics[scale=0.7]{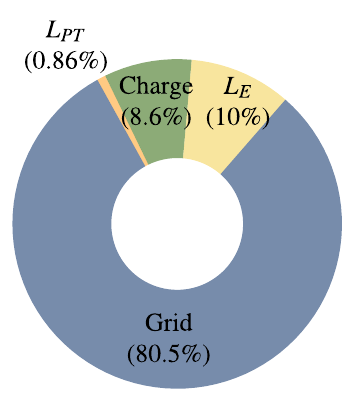}
    \caption{Generator energy.}
    \label{subfig:Pie_Gen_C_3}
    \end{subfigure}%
    \begin{subfigure}{0.5\columnwidth}
    \centering
    \includegraphics[scale=0.7]{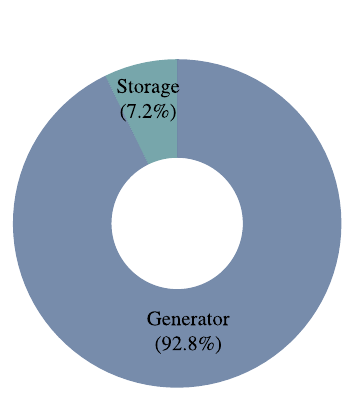}
    \caption{Revenue.}
    \label{subfig:Pie_rev_C_3}
    \end{subfigure}%
    \captionsetup[figure]{justification=centering}
    \caption{Breakdown of various elements in Case Study \rom{3}, a nuclear power plant with hydrogen generation and storage facility: (a) Generator energy usage, and (b) Revenue contributions.
    }
    \label{fig:Pie_C_3}
\end{figure}

The optimal storage capacity is $195930~[\unit{kg}]$, with the NPV objective function of $\$-6.27\times 10^{9}$, indicating that under current assumptions, including retail prices, facility costs, and lifetime of the plant, the project needs further time to become profitable.
However, since the optimal storage capacity is nonzero, storage improves the economics of the new plant.
Figure~\ref{subfig:Pie_Gen_C_3} indicates that from the total generator's energy, $80.5\%$ was directly sold to the electricity grid, while $8.6\%$ was used for charging the tertiary hydrogen storage, $10\%$ used to satisfy auxiliary electrical loads, and $0.86\%$ used to satisfy the primary load demand for HTSE. 
In this case study and with the presented assumptions, the storage is responsible for $7.2\%$ of the generated revenue over the lifetime of the project.
This result is shown in Fig.~\ref{subfig:Pie_rev_C_3}.

According to Ref.~\cite{MACKERRON1992641}, $60\%$ to $70\%$ of the cost of power produced by NPPs is directly related to capital costs.
The addition of a hydrogen storage facility, while improving flexibility, increases costs, for example by $37\%$ in Ref.~\cite{rad2020techno}.  
Therefore, 
the consideration of potential uncertainties in overnight construction costs ($C_{\text{occ}}$) can inform project management, planning, and scheduling.
It can also provide unique opportunities for understanding risks \textemdash{} informing further actions to protect investors \cite{Wealer2021}.
The market prices for hydrogen, which are also inherently uncertain, have a significant impact on optimal system behavior and its economic performance.
The impact of these uncertainties on the optimal performance of an NPP with a hydrogen production and storage facility is explored next.

Here, we assume that the OCC associated with hydrogen storage is an uncertain variable defined within  $C_\text{occ,Hyd} \in [400~800]~[\unit{\$/kW}]$.
In addition, due to the significant impact of hydrogen market prices on optimal system behavior and its inherent uncertainties, $C_{T} \in [3~10]~[\unit{\$/k}]$ is also assumed to be an uncertain variable.  
In the following, we assess the impact of variations in $C_\text{occ}$ and hydrogen market prices on the optimal HES solution. 
Informed by the lack of data to construct the probability distribution, a crisp representation of uncertainties is used to characterize these uncertain quantities \cite{Azad2023}.
Various levels of overnight construction costs are then employed, along with various levels of hydrogen market prices, to construct surfaces for optimal NPV and storage capacity.

\begin{figure}[t]
    \captionsetup[subfigure]{justification=centering}
    \centering
    \begin{subfigure}{0.5\columnwidth}
    \centering
    \includegraphics[scale=0.453]{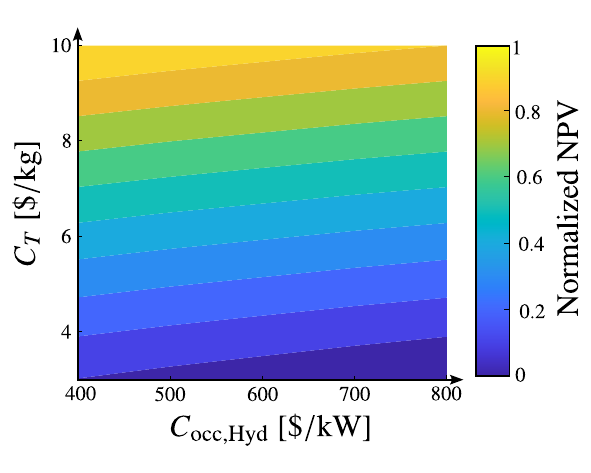}
    \caption{Optimal NPV.}
    \label{subfig:NPP_OCC_NPV}
    \end{subfigure}%
    \begin{subfigure}{0.5\columnwidth}
    \centering
    \includegraphics[scale=0.453]{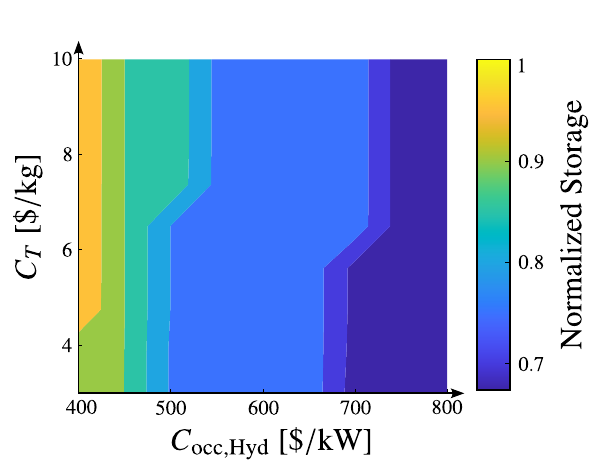}
    \caption{Optimal Storage.}
    \label{subfig:NPP_OCC_P}
    \end{subfigure}%
    \captionsetup[figure]{justification=centering}
    \caption{Case Study \rom{3}: Optimal NPV and storage capacity for uncertainties in overnight construction cost of nuclear power plant and hydrogen market prices.}
    \label{fig:uncertain}
\end{figure}

The results from these investigations are shown in Fig.~\ref{fig:uncertain}. 
Specifically, Figs.~\ref{subfig:NPP_OCC_NPV} and \ref{subfig:NPP_OCC_P} show variations in normalized optimal NPV and storage capacity when variations occur in both $C_{\text{occ,Hyd}}$ and $C_{T}$.
From these results, it is clear that in the presence of uncertainties from OCC of the hydrogen storage and hydrogen market prices, the maximum NPV corresponds to the case with the lowest $C_\text{occ,Hyd}$ and maximum $C_{T}$.
This conclusion is shown in Fig.~\ref{subfig:NPP_OCC_NPV}.
As shown in Fig.~\ref{subfig:NPP_OCC_P}, under the assumption of a prescribed generator capacity, and associated price signals, the optimal storage size is also reduced with high OCC and low hydrogen prices.

\xsection{Conclusion}\label{sec:conclusion}

In this article, we developed an efficient framework to assess the economic feasibility of integrated generator and storage energy systems.
The proposed framework, referred to as \ToolName, empowers the early-stage investigations of various concepts and configurations of integrated energy systems.
It leverages a system-level, control co-design approach to identify the most profitable technology parameters and operations under specific assumptions. 
\ToolName~has the potential to offer unique insights for both new projects and retrofit efforts, facilitating the decision-making process and communications among technology experts, investors, and stakeholders.

The capabilities of \ToolName~were tested for three case studies.
Case Study \rom{1} was focused on a power plant that includes a natural gas combined cycle generator, a thermal storage unit, and a carbon capture and storage system.
Case Study \rom{2} considered a wind farm with a battery storage system, thereby bringing up some of the complexities in dealing with non-dispatchable renewable technologies.
Hybrid operation of a nuclear power plant with a hydrogen generation and storage facility was assessed in Case Study \rom{3}. 
Since in its current form, \ToolName~solves a linear optimization problem, the case studies were investigated for the duration of $30$ years, on an hourly basis, in an efficient manner, identifying the global optimal solution.

Since the proposed framework relies on simplified, fundamental subsystem dynamics based on basic energy balance and operating states, a potential user may find that these model assumptions are inadequate for a selected system (e.g., temperature dependence when using an NGCC or more detailed physics associated with the HTSE facility).
Their dynamics (and other constraints) may then be updated with other differential-algebraic equations to fit within the general framework, but this may result in potential trade-offs in the efficiency of the optimization and analysis that can be performed.
However, \ToolName{} provides an initial foundation with a relevant basic fidelity level that a user can build upon.%

As a next step, it is desirable to enable the framework to account for the simultaneous presence of multiple storage types and other heterogeneous and more complex configurations.
This framework, within its current and future capabilities, has the potential to further improve the economic viability of various integrated energy systems and the decision-making surrounding them.

\begin{acknowledgment}
The authors would like to thank Ian Goven and Roberto Vercellino for their contributions to the development of this research.
\end{acknowledgment}

\renewcommand{\refname}{REFERENCES}
\bibliographystyle{asmems4}
\begin{mySmall}
\bibliography{References}
\end{mySmall}

\appendix
\xsection{Appendix}\label{sec:Appendix}
This section includes some complementary information, providing more details for understanding the optimization model. 

\subsection{Node Definitions}\label{sec:Appendix_nodes}
Mathematical descriptions of the nodes labeled in Fig.~\ref{fig:IES} are presented in this section. 
These equations, which are used to formulate inequality constraints associated with Eqs.~(\ref{Eq:n146})--(\ref{eq:LGPLGE}), present the available amount of power from the generator at different stages, are described in further detail in Tab.~\ref{Tab:SignalTable}.

\newlength{\mywidth}
\setlength{\mywidth}{9.8cm}

\newlength{\myspace}
\setlength{\myspace}{3pt}

\renewcommand{\tabularxcolumn}[1]{>{\arraybackslash}m{#1}}

\begin{table*}[b]
\centering
\caption{Node signal definitions with lexical interpretation from Fig.~\ref{fig:IES}.}
\label{Tab:SignalTable}

\begin{tabularx}{1\textwidth}{ X  m{\mywidth}  }   
\hline \hline
\textbf{Lexical interpretation} 
& \textbf{Mathematical description} \\ 

\hline 
Initial available primary-type power 
&
${n_{1}}: {x_{G}}$ 
\\      

\cline{1-2}
Available primary-type power after charging primary storage units
&
${n_{2}}: {x_{G} - \overrightharpoonsub{u}{P}}$
\\

\cline{1-2}
Available primary-type power after satisfying primary loads
&
${n_{3}}: {x_{G} - \overrightharpoonsub{u}{P} - L_{GP}}$
\\

\cline{1-2}
Available primary-type power after satisfying the primary load for tertiary operation
& 
${n_{4}}:{x_{G} - \overrightharpoonsub{u}{P} - L_{GP} - L_{GPT}}$
\\

\cline{1-2}
Available electrical power after accounting for power conversion efficiency
&
${n_{5}}: \eta_{G}(x_{G} - \overrightharpoonsub{u}{P} - L_{GP} - L_{GPT})$
\\

\cline{1-2}
Available electrical power after charging electrical storage
&
${n_{6}}: {\eta_{G}(x_{G} - \overrightharpoonsub{u}{P} - L_{GP} - L_{GPT})- \overrightharpoonsub{u}{E}}$
\\

\cline{1-2}
Available electrical power after satisfying electrical load
&
${n_{7}}: {\eta_{G}(x_{G} - \overrightharpoonsub{u}{P} - L_{GP} - L_{GPT})- \overrightharpoonsub{u}{E} - L_{GE}}$
\\ 

\cline{1-2}
Available electrical power after satisfying electrical load for tertiary operation
&
${n_{8}}:{\eta_{G}(x_{G} - \overrightharpoonsub{u}{P} - L_{GP} - L_{GPT} )- \overrightharpoonsub{u}{E} - L_{GE} - \overrightharpoonsub{u}{T}}$
\\

\cline{1-2}
Available electrical power after combusting tertiary product
&
${n_{9}}: {\eta_{G}(x_{G}} - {\overrightharpoonsub{u}{P}} - {L_{GP} - L_{GPT})} - {\overrightharpoonsub{u}{E} - L_{GE}} - {\overrightharpoonsub{u}{T}} + {\alpha_{c}\overrightharpoonsub{\eta}{T}\overrightharpoonsub{u}{T}} - {\alpha_{c}\overrightharpoonsub{\eta}{T}u_{TR}}$
\\

\cline{1-2}
Primary-type power supplied by the generator to satisfy the primary load demand
& $L_{GP}  = L_{P}x_{G} - \overrightharpoonsub{\eta}{P}\overrightharpoonsub{u}{P} + \overrightharpoonsub{\eta}{P}u_{PR}$
\\

\cline{1-2}
Primary-type power supplied by the generator to satisfy the primary load demand for tertiary operation
& $L_{GPT}  = L_{PT}\overrightharpoonsub{u}{T}$
\\

\cline{1-2} 
Electrical power supplied by the generator to satisfy the electrical load demand
& $L_{GE}  = L_{E}x_{G} - \overrightharpoonsub{\eta}{E}\overrightharpoonsub{u}{E} + \overrightharpoonsub{\eta}{E}u_{ER}$
\\

\hline \hline
\end{tabularx}
\end{table*}

\subsection{Optimization Model}\label{sec:Appendix_optmodel}
\begin{figure*}[b]
    \centering
    \includegraphics[scale = 0.60]{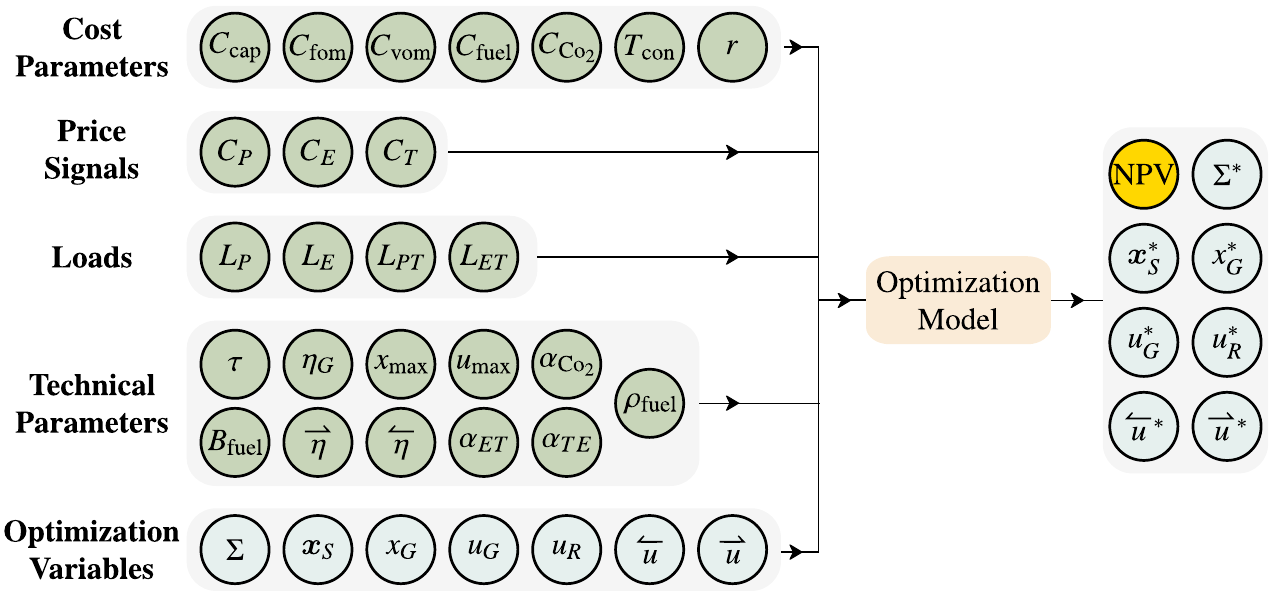}
    \caption{Optimization model with inputs and outputs.} 
    \label{fig:OptModel}
\end{figure*}

The inputs into the optimization model, along with outputs consisting of the NPV objective function and optimized decision variables, are graphically illustrated in Fig.~\ref{fig:OptModel}. 
Note that this description is representative, and depending on the problem at hand, more advanced technical and economic parameters may be included.

\subsection{Lexical Interpretations of Problem Elements}\label{sec:Appendix_lex}
This section serves as a complementary section to the mathematical explanations offered throughout the article.
Specifically, in this section, we offer some lexical interpretations to facilitate the understanding of various problem elements by non-optimization experts.  
These interpretations are offered in Tab.~\ref{Tab:lexicalInterp}.

\nomenclature[A, 01]{\(\textrm{BESS}\)}{battery energy storage system}
\nomenclature[A, 02]{\(\textrm{CCD}\)}{control co-design}
\nomenclature[A, 03]{\(\textrm{CCS}\)}{carbon capture and storage}
\nomenclature[A, 04]{\(\textrm{HES}\)}{hybrid energy systems}
\nomenclature[A, 05]{\(\textrm{HTSE}\)}{high-temperature steam electrolysis}
\nomenclature[A, 07]{\(\textrm{LWR}\)}{light water reactor}
\nomenclature[A, 08]{\(\textrm{NGCC}\)}{natural gas combined cycle}
\nomenclature[A, 09]{\(\textrm{NPV}\)}{net present value}
\nomenclature[A, 10]{\(\textrm{NPP}\)}{nuclear power plant}
\nomenclature[A, 11]{\(\textrm{TES}\)}{thermal energy storage}
\nomenclature[A, 12]{\(\textrm{OCC}\)}{overnight capital cost}

\nomenclature[B, 01]{\( \parm_{E}  \)}{index for electrical domain}
\nomenclature[B, 02]{\( \parm_{G}  \)}{index for generator}
\nomenclature[B, 03]{\( \parm_{P}  \)}{index for primary domain}
\nomenclature[B, 0r]{\( \parm_{R}  \)}{index for revenue}
\nomenclature[B, 05]{\( \parm_{S}  \)}{index for storage}
\nomenclature[B, 06]{\( \parm_{T}  \)}{index for tertiary domain}

\nomenclature[v, 01]{\( C_{\text{cap}}  \)}{capital cost}
\nomenclature[v, 02]{\( \text{CO\textsubscript{2}}  \)}{CO\textsubscript{2} penalty}
\nomenclature[v, 03]{\( C_{E}  \)}{electricity price signal}
\nomenclature[v, 04]{\( C_{\text{fom}}  \)}{fixed O\&M cost}
\nomenclature[v, 05]{\( C_{\text{fuel}}  \)}{fuel price signal}
\nomenclature[v, 06]{\( C_{P}  \)}{primary price signal}
\nomenclature[v, 07]{\( C_{T}  \)}{tertiary price signal}
\nomenclature[v, 08]{\( C_{\text{vom}}  \)}{variable O\&M cost}
\nomenclature[v, 09]{\( E_{\text{fuel}}  \)}{fuel expenditure}
\nomenclature[v, 10]{\( L \)}{load}
\nomenclature[v, 11]{\( n  \)}{node}
\nomenclature[v, 12]{\( R  \)}{revenue}
\nomenclature[v, 13]{\( r  \)}{discount rate}
\nomenclature[v, 14]{\( T_{\text{con}}  \)}{construction time}
\nomenclature[v, 15]{\( \bm{u}  \)}{control vector}
\nomenclature[v, 16]{\( \overrightharpoon{u}  \)}{charging signal}
\nomenclature[v, 17]{\( \overleftharpoon{u}  \)}{discharging signal}
\nomenclature[v, 18]{\( \bm{x}  \)}{state vector}
\nomenclature[v, 19]{\( \bm{\Sigma}  \)}{storage capacity}
\nomenclature[v, 20]{\( \overrightharpoon{\eta}  \)}{charge efficiency}
\nomenclature[v, 21]{\( \overleftharpoon{\eta}  \)}{discharge efficiency}
\nomenclature[v,22]{\( \alpha_{\parm} \)}{conversion rate}

\printnomenclature


\setlength{\mywidth}{5cm}
\setlength{\myspace}{3pt}

\renewcommand{\tabularxcolumn}[1]{>{\arraybackslash}m{#1}}

\begin{table*}[t]
\centering
\caption{Lexical interpretation for select elements in the optimization problem.}
\label{Tab:lexicalInterp}

\begin{tabularx}{1\textwidth}{ X  m{\mywidth}  m{2.5cm} }  
\hline \hline
\textbf{Lexical interpretation} 
& \textbf{Mathematical description} & \textbf{Reference} \\

\hline
The objective function is the \GS{maximization} of the \OR{Net Present Value} of the system over the lifetime of the power plant 
&
\GS{maximize} \OR{NPV} 
& Eq.~(\ref{eq:NPV})
\\

\hline
\GS{Generator state dynamics} describes the power level of the generator unit as a function of \OR{ramp rate}, \GS{generator's current state}, and the \GS{requested power}
&
\GS{$\dot{x}_{G}(t)$} = $\frac{1}{\OR{\tau}}(-\GS{x_{G}(t)} + \GS{u_{G}(t)} )$
&
Eq.~(\ref{eq:generatordynamics})
\\

\hline
\GS{Storage state dynamics} describes the available resource in the storage system as a function of \GS{charging} and \GS{discharging} power signals, and the storage \OR{efficiencies}
& $\GS{\dot{\bm{x}}_{S}(t)} = \OR{\overrightharpoon{\eta}}\GS{\overrightharpoon{\bm{u}}(t)} - \OR{\overleftharpoon{\eta}}\GS{\overleftharpoon{\bm{u}}(t)}  $
&
Eq.~(\ref{eq:Storagedynamics})
\\
\hline
\GS{Storage capacity}  must be  \CR{non-negative}
&
$\CR{\bm{0}} \leq \GS{\bm{\Sigma}}$   
&
Eq.~(\ref{eq:plantcons1})
\\

\hline
At every time instant, \GS{control variables} are \CR{non-negative} and limited by a \OR{maximum limit}
&
$\CR{\bm{0}} \leq \GS{\bm{u}(t)} \leq \OR{\bm{u}_{\text{max}}}$ 
&
Eq.~(\ref{eq:controlcons1})
\\

\hline
\GS{Revenue control signal} is never greater than the control \OR{discharged power}
&
\begin{minipage}{\mywidth}
\vspace{\myspace}
$\GS{u_{PR}(t)}  \leq \OR{\overleftharpoonsub{u}{P}(t)}$ \\
$\GS{u_{ER}(t)}  \leq \OR{\overleftharpoonsub{u}{E}(t)}$  \\
$\GS{u_{TR}(t)}  \leq \OR{\overleftharpoonsub{u}{T}(t)}$
\vspace{\myspace}
\end{minipage}
&
Eqs.~(\ref{Eq:P_rev})--(\ref{Eq:T_rev})
\\

\hline
The \GS{generator's power level} is \CR{non-negative} and limited by \OR{nominal capacity} or \OR{maximum available power to extract}
& 
$\CR{\bm{x}_{G,\text{min}}(t)} \leq \GS{\bm{x}_{G}(t)} \leq \OR{\bm{x}_{G,\text{max}}(t)}$
&
Eq.~(\ref{eq:controlcons2})
\\

\hline    
The \GS{storage energy level} is \CR{non-negative} and never greater than the \OR{storage capacity} 
&
$\CR{\bm{0}} \leq \GS{\bm{x}_{S}(t)} \leq \OR{\bm{\Sigma}}$
&
Eq.~(\ref{eq:storagecons})
\\

\hline 
The \GS{generator} and \GS{storage} states are prescribed \OR{a specific value at $t_{0}$} and storage may also be prescribed \CR{to the initial value at $t_{f}$}
&
\begin{minipage}{\mywidth}
$\GS{\bm{x}(t_{0})} = \OR{\bm{x}_{0}}$ \\ 
$\CR{\bm{x}_{S}(t_{f})} = \OR{\bm{x}_{S}(t_{0})}$
\end{minipage}
&
Eq.~(\ref{Eq:initfinal})
\\

\hline 
Each \GS{charging signal} is limited by the \OR{available power in the generator} at that node
&
\begin{minipage}{\mywidth}
\vspace{\myspace}
$\GS{\overrightharpoonsub{u}{p}(t)}  \leq  \OR{n_{1}(t)}$ \\
$\GS{\overrightharpoonsub{u}{E}(t)}  \leq  \OR{n_{5}(t)}$ \\
$\GS{\overrightharpoonsub{u}{T}(t)}  \leq  \OR{n_{7}(t)}$
\vspace{\myspace}
\end{minipage}
&
Eq.~(\ref{Eq:n146})
\\
\hline 
The \GS{generator's power signals to satisfy primary and electrical loads} is \CR{non-negative} and always smaller or equal to the \OR {available power in the corresponding node}
&
\begin{minipage}[b]{\mywidth}
 \vspace{\myspace}
$\CR{\bm{0}} \leq \GS{ L_{GP}(t)} \leq \OR{n_{2}(t)}$ \\
$\CR{\bm{0}} \leq \GS{ L_{GPT}(t)} \leq \OR{n_{3}(t)}$ \\
$\CR{\bm{0}} \leq \GS{ L_{GE}(t)} \leq \OR{n_{5}(t)}$
\end{minipage}
&
Eq.~(\ref{eq:LGPLGE})
\\
\hline \hline
\end{tabularx}
\end{table*}

\end{document}